\documentclass{mn2e}
\input psfig.sty

\newcommand{\ee}{$e^\pm$}
\newcommand{\ef}{E_{\rm f}}
\newcommand{\ep}{E_{\rm p}}
\newcommand{\g}{$\gamma$}

\newcommand{\sax}{{\it Beppo\-SAX}}

\newcommand{\xte}{{\it RXTE}}

\newcommand{\ginga}{{\it Ginga}}
\newcommand{\asca}{{\it ASCA}}
\newcommand{\gro}{{\it CGRO}}
\newcommand{\cnu}{\chi^2/\nu}
\newcommand{\wfe}{W_{\rm K\alpha}}

\title{Correlations between X-ray and radio spectral properties of accreting
black holes}

\author[A. A. Zdziarski et al.]
{Andrzej A. Zdziarski,$^1$ Piotr Lubi\'nski,$^1$ Marat Gilfanov$^{2,3}$ and Mike
Revnivtsev$^3$\\
$^1$N. Copernicus Astronomical Center, Bartycka 18, 00-716 Warsaw, Poland\\
$^2$Max-Planck-Institut f\"ur Astrophysik, Karl-Schwarzschild-Str. 1, 85740
Garching, Germany\\
$^3$Space Research Institute, Russian Academy of Sciences, Profsoyuznaya 84/32,
117810 Moscow, Russia\\
}

\date{Accepted 2003 February 20. Submitted 2002 September 18}

\pagerange{\pageref{firstpage}--\pageref{lastpage}}
\pubyear{2003}

\begin{document}

\maketitle

\label{firstpage}

\begin{abstract} We study correlations betwen X-ray spectral index, strength of
Compton reflection, and X-ray and radio fluxes in accreting black holes
(Seyferts and black-hole binaries). We critically evaluate the evidence for the
correlation of the X-ray spectral index with the strength of Compton
reflection and consider in detail statistical and systematic effects that can
affect it. We study patterns of spectral variability (in particular, pivoting of
a power law spectrum) corresponding to the X-ray index-flux correlation. We also
consider implications of the form of observed X-ray spectra and their
variability for interpretation of the correlation between the radio and X-ray
fluxes. Finally, we discuss accretion geometries that can account for the
correlations and their overall theoretical interpretations. \end{abstract}


\section{INTRODUCTION}
\label{intro}

X-ray and soft \g-ray (hereafter X$\gamma$) spectra from luminous accreting
black holes (hereafter BH), i.e., AGNs and BH binaries, commonly show a distinct
component due to Compton reflection (Lightman \& White 1988; Magdziarz \&
Zdziarski 1995) of the primary continuum from a cold medium (e.g., Pounds et
al.\ 1990; Nandra \& Pounds 1994; Magdziarz et al.\ 1998; Weaver, Krolik \& Pier
1998; Zdziarski, Lubi\'nski \& Smith 1999, hereafter ZLS99; Done, Madejski \&
\.Zycki 2000; Eracleous, Sambruna \& Mushotzky 2000; Done et al.\ 1992;
Gierli\'nski et al.\ 1997; Zdziarski et al.\ 1998; \.Zycki, Done, \& Smith 1998;
1999; Gilfanov, Churazov \& Revnivtsev 1999, 2000, hereafter GCR00; Revnivtsev,
Gilfanov \& Churazov 1999, 2001). A very interesting property of Compton
reflection with a number of potential physical implications is that its relative
strength, $\sim \Omega/2\upi$, where $\Omega$ is the solid angle of the cold
reflector as seen from the hot plasma, correlates with some other spectral and
timing properties of many sources (e.g., ZLS99; GCR00).

Another correlation often found in both Seyferts and BH binaries is between 
X-ray spectral index and X-ray flux (e.g., Chiang et al.\ 2000; Done et al.\ 
2000; Nowak, Wilms \& Dove 2002; Zdziarski et al.\ 2002b, hereafter Z02; Lamer 
et al.\ 2003; Gliozzi, Sambruna \& Eracleous 2003). The X-ray flux is also 
correlated with the level of radio emission in the hard states of BH binaries 
(Corbel et al.\ 2000; Gallo, Fender \& Pooley 2003). All these correlations 
appear to reflect fundamental properties of BH accretion flows. We critically 
study the correlations, relationships between them, their theoretical models, 
and the corresponding physical implications. In Appendix A, we present 
properties of spectral variability due to a power-law pivoting, which process is 
closely related to the flux-index correlation as well as it puts constraints on 
the interpretation of the radio--X-ray correlation. In spectral fits, we use 
{\sc xspec} (Arnaud 1996).

\section{Correlation of Compton reflection with spectral index}
\label{ref_index}

\subsection{Results of fits to data}
\label{data}

The first to find a correlation of Compton reflection with another spectral
property were Ueda, Ebisawa \& Done (1994), who found that $\Omega$ correlates
with the X-ray photon spectral index, $\Gamma$, in the BH candidate GX\,339--4,
albeit their result was based on only five observations (by \ginga). Later, 23
\ginga\/ observations of BH and neutron-star binaries were found to obey the
same correlation (Zdziarski 1999).

\begin{figure}
\centerline{\psfig{file=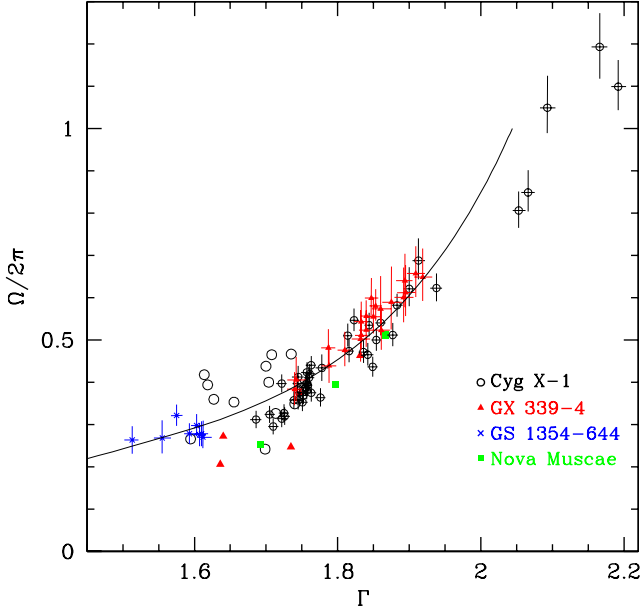,width=8.5cm}}
\caption{Correlation between the strength of Compton reflection and the
X-ray spectral index in BH binaries. Small symbols with error bars and
large ones without them correspond to observations by \xte\/ (GCR00 and
references therein) and \ginga\/ (Zdziarski 1999 and references therein),
respectively. The solid curve corresponds to a model of ZLS99
with a central hot source surrounded by an overlapping cold disc.
\label{og_bhb} }
\end{figure}

\begin{figure}
\centerline{\psfig{file=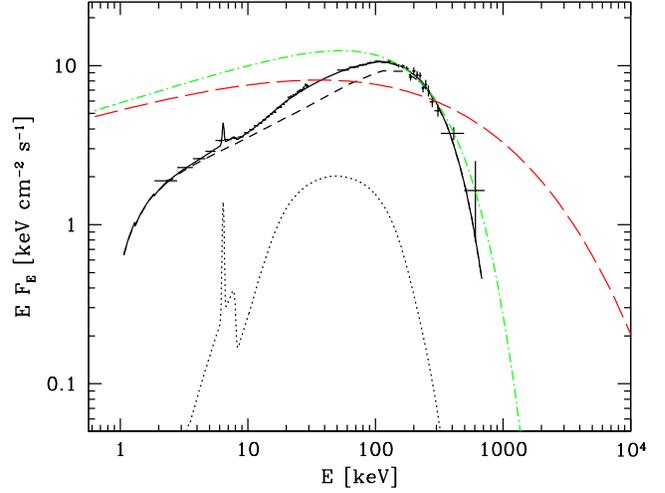,width=8.5cm}}
\caption{A broad-band $\sim 1$--$10^3$ keV spectrum of Cyg X-1 (Gierli\'nski et
al.\ 1997) fitted by thermal Comptonization and Compton reflection (black solid
curve). The two components separately are shown by the short dashes and dots, respectively. At low energies, the spectrum is absorbed in interstellar and
circumstellar media. The long dashes and dot/dashes show (unabsorbed) nonthermal
synchrotron spectra emitted by power-law electrons with $p=2.5$ and with an
exponential and sharp high-energy cutoff, respectively (see Section \ref{radio}).
\label{cygx1} }
\end{figure}

The reality of the correlation has been unambiguously confirmed in the \xte\/
data for the luminous BH binaries Cyg\,X-1, GX\,339--4 and GS\,1354--644
(Gilfanov et al.\ 1999; GCR00; Revnivtsev et al.\ 2001). Fig.\ \ref{og_bhb}
presents those \xte\/ results, as well as the \ginga\/ results for 20
observations of Cyg\,X-1, GX\,339--4 and Nova Muscae. These results were mostly
obtained with spectra constrained to energies below a few tens of keV.
However, in many instances, we also have at our disposal broad-band spectra
extending to several hundred of keV, where almost the entire range of the
reflection spectrum is found to fit the data very well, together with thermal
Comptonization (e.g., the hard states of Cyg X-1, Gierli\'nski et al.\ 1997, and
GX 339--4, Zdziarski et al.\ 1998), as illustrated in Fig.\ \ref{cygx1}.

The $\Omega$-$\Gamma$ correlation is also seen in Fourier-resolved spectra,
i.e., corresponding to variability in a given range of Fourier frequencies
(Revnivtsev et al.\ 1999, 2001; Gilfanov et al.\ 1999). Very importantly, the
strength of reflection was also found to correlate with the low-frequency QPO
centroid frequency and with the degree of the relativistic smearing of the Fe
K$\alpha$ line (associated with reflection, \.Zycki \& Czerny 1994), see
fig.\ 4 in Gilfanov et al.\ (1999) figs.\ 1.4 and 1.6 in GCR00, fig.\ 2 in
Revnivtsev et al.\ (2001), and Gilfanov et al., in preparation.

We note that the above results concern BH binaries in their luminous (and mostly
hard) states. Existing data are insufficient to constrain reflection in
quiescent states of BH binaries. However, given the prevalent theoretical
interpretation of the correlation as due to feedback between a cold accretion
disc and hot plasma (Section \ref{feedback}), we do not expect it to be present
in quiescence as the disc is then cut off at a large radius (e.g., Narayan \& Yi
1995) and Comptonization of synchrotron photons usually dominates X-rays (e.g.,
Wardzi\'nski \& Zdziarski 2000). Analogously, we do not expect an
$\Omega$-$\Gamma$ correlation in low-luminosity AGNs.

The picture of the correlation disappearing with the decreasing luminosity is 
indeed confirmed in the BH transient GS 2023+338. The results of \.Zycki et al.\ 
(1999) show an $\Omega$-$\Gamma$ correlation in the initial luminous hard state, 
but then the spectrum substantially softened at an approximately constant 
$\Omega$ when the luminosity decreased by a factor $\ga 10$. This effect is most 
likely due to the onset of the dominance of Comptonization of synchrotron 
photons with decreasing luminosity (Wardzi\'nski \& Zdziarski 2000). Similarly, 
reflection is weak in the transient XTE J1118+480 (Frontera et al.\ 2001b; 
Miller et al.\ 2002), which maximum luminosity was only about $\sim 10^{-3}$ of 
the Eddington one\footnote{An additional complicating factor in measuring 
reflection in this halo system is its likely overall low metallicity, which 
possibility was considered by Frontera et al.\ (2001b), but not by Miller et 
al.\ (2002), who only considered the case of a low Fe abundance at all other 
abundances kept solar. Thus, the actual value of $\Omega/2\upi$ in this object 
may still be $\sim 0.2$ found by the former authors.}.

\begin{figure}
\centerline{\psfig{file=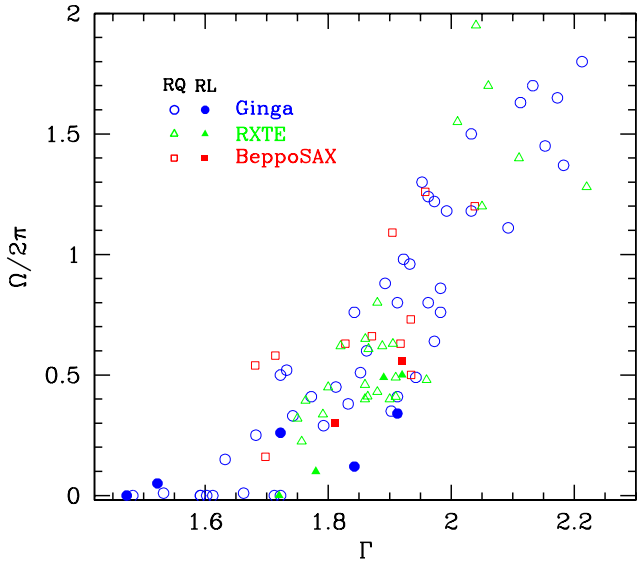,width=8.5cm}}
\caption{Correlation between the strength of Compton reflection and the
X-ray spectral index (in the $\sim 3$--20 keV range, see Section
\ref{systematic} and eq.\ [\ref{delta}]) in AGNs.
Open and filled symbols correspond to radio-quiet AGNs and active radio
galaxies, respectively. Blue circles, green triangles, and red squares
correspond to observations by \ginga, \xte, and \sax, respectively. Error
bars/contours are not shown for clarity, see Fig.\ \ref{contours}a for \ginga\/
error contours. References for \ginga: Lubi\'nski \& Zdziarski (2001), Wo\'zniak
et al.\ (1998); \xte: Weaver et al.\ (1998), Lee et al.\ (1998), Done et al.\
(2000), Chiang et al.\ (2000), P02a, Eracleous et al.\ (2000), this work (NGC
7469 in Section \ref{samples}); \sax: P02b, Orr et al.\ (2001), ZG01, Grandi et
al.\ (2001).
\label{og_agn} }
\end{figure}

The first to report an $\Omega$-$\Gamma$ correlation in an AGN were Magdziarz et
al.\ (1998), in the Seyfert-1 galaxy NGC 5548. Then, ZLS99 showed the presence
of a strong $\Omega$-$\Gamma$ correlation in 47 \ginga\/ observations of 23 AGNs
(mostly Seyfert 1s and a few intermediate-type AGNs). Matt (2001) presented a
compilation of \sax\/ results for Seyfert 1s, and found that $\Omega$ is
correlated with $\Gamma$ in the full sample at the confidence level of $>0.999$,
and $\sim 0.99$ if one object in which the measurement of the continuum was
probably affected by a soft excess is not taken into account. A similar sample
was later studied by Perola et al.\ (2002, hereafter P02b). Zdziarski \& Grandi
(2001, hereafter ZG01) showed that Compton reflection in broad-line radio
galaxies (also called radio-loud Seyfert 1s), albeit weaker on average than in
radio-quiet AGNs, is still consistent with the same correlation. Nandra et al.\
(2000, hereafter N00) found an $\Omega$-$\Gamma$ correlation in multiple \xte\/
observations of the Seyfert galaxy NGC 7469. Papadakis et al.\ (2002, hereafter
P02a) have found this correlation in the average properties of four Seyferts.
Fig.\ \ref{og_agn} summarizes those (and some other) results.

It is, however, of importance to consider possible effects that may lead to
spurious $\Omega$-$\Gamma$ correlations. We consider below in detail
statistical and systematic effects.

\subsection{Statistical effects}
\label{statistical}

The main statistical effect is due to the fitted strength of reflection and
spectral index for a single observation being correlated to certain degree,
resulting in a skewness of their joint error contour. Any data have limited
statistics, and then the same intrinsic $\sim 2$--20 keV spectrum can be fitted
within some confidence limit with either a somewhat harder index and less
reflection or a softer index and more reflection. This effect is more important
for AGNs than for BH binaries, as the former have usually significantly lower
statistics.

\begin{figure}
\centerline{\psfig{file=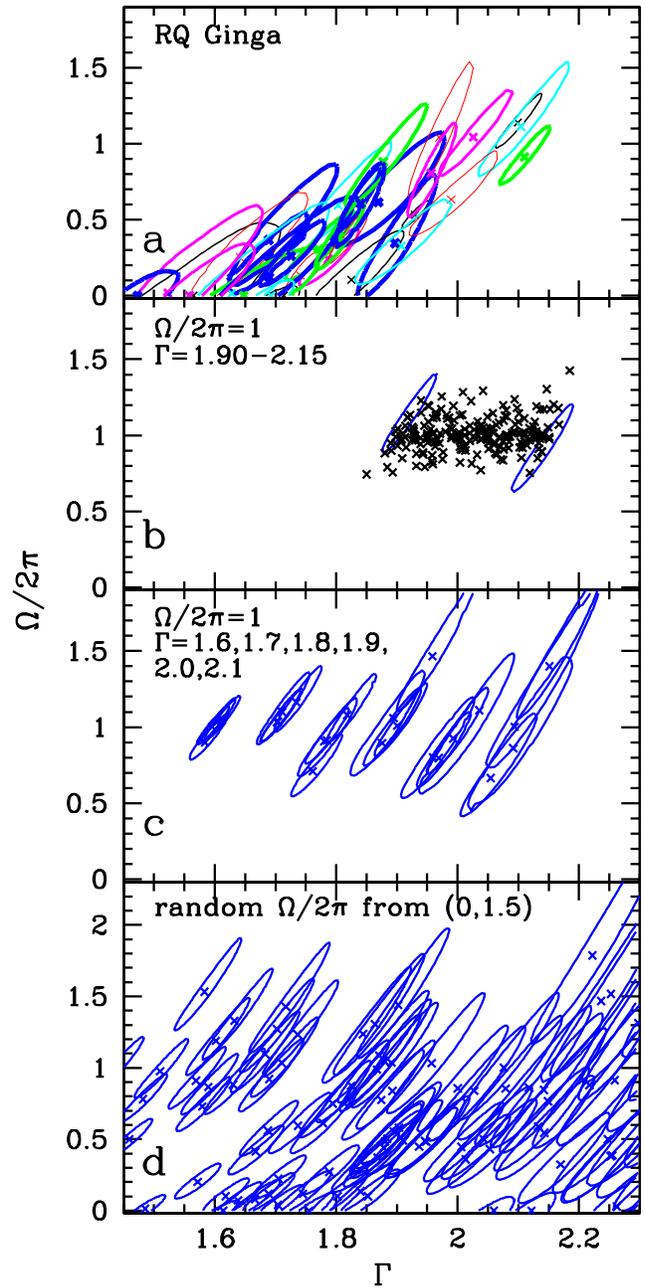,width=8.5cm}}
\caption{(a) Error contours ($1\sigma$ for 2 parameters) for the correlation
between $\Omega$ and $\Gamma$ in the \ginga\/ data of ZLS99 for AGNs. For
clarity of display, only contours with the vertical extent of
$\Delta(\Omega/2\upi)<1$ are shown, and alternating colors
and line widths are used. The panels below show simulations, see text for
details. (b) Best fits and 2 representative contours for
simulated data with a uniform distribution of $\Gamma$ and fixed
$\Omega$. (c) Fixed $\Omega$ for 6 values of $\Gamma$. (d) Random $\Omega$ for a
uniform distribution of $\Gamma$. Clearly, all simulations results are very
different from the observed distribution.
\label{contours} }
\end{figure}

As usual with this type of effects, its importance decreases with increasing
statistics. As shown in Fig.\ \ref{contours}a, the extent of typical
error contours for Seyferts is {\it much less\/} than the global extent of the
correlation, and the inclination of an individual contour is also significantly
steeper from that of the correlation. Thus, it is already very unlikely that
statistical effects alone can account for this distribution.

Then, Vaughan \& Edelson (2001, hereafter VE01) considered 120 simulated \xte\/
spectra based on observations of MCG --6-30-15 with short exposure times of
$\sim 1.4$--3.7 ks and the top layer of the \xte\/ PCUs 0--2. They used a
power-law model with $\Gamma$ in a relatively narrow range, $\sim 1.9$--2.15 for
90 per cent of the simulated spectra, and constant reflection at
$\Omega/2\upi=1.42$. They fitted the simulated spectra with $\Gamma$ and
$\Omega/2\upi$ as free parameters. The fitted values of $(\Gamma,\,
\Omega/2\upi)$ formed a skewed elongated contour extending from $\sim
(1.75,\,0.5)$ to $\sim (2.3,\,3)$.

Given the assumed narrow range of $\Gamma$, the $(\Gamma,\,\Omega/2\upi)$
contour obtained by them is equivalent (except for some spread in the values of
$\Gamma$ and the exposure time) to the statistical error contour for two
parameters. Indeed, the typical error bars provided by VE01 are comparable to
the extent of their contour. However, based on the extent of this contour, VE01
stated that the above statistical effect casted serious doubts on the claim
of the correlation by ZLS99.

However, most of the error contours of ZLS99 (Fig.\ \ref{contours}a) and of
those corresponding to \xte\/ and \sax\/ results of Fig.\ \ref{og_agn} are much
smaller than the contour obtained by VE01 due to much better statistisc. In
particular, the typical statistics in the \ginga\/ spectra is an order of
magnitude better than those of the 1-orbit simulated spectra of VE01. Also, the
shape of the error contour of VE01 is significantly different from those
observed (Fig.\ \ref{og_agn}). Already based on the above, the suggestion of
VE01 is unlikely to be correct.

In order to quantitatively study this issue, we have generated 208 simulated 
\xte\/ spectra (assuming a 0.5 per cent systematic error, which we estimated 
from our fits to Crab data) with the range of $\Gamma$ of 1.90--2.15 (similar to 
that of VE01) and the constant $\Omega/2\upi=1$. In the simulations, we have 
assumed a source with the constant 2--10 keV flux of 3.5 (flux unit is 
$10^{-11}$ erg cm$^{-2}$ s$^{-1}$ hereafter in this paragraph) and the exposure 
of 50 ks for PCUs 0--2 and 30 ks for PCUs 3--4, similar to that of observations 
of NGC 7469, see Section \ref{samples} below. The product of the flux and 
exposure roughly corresponds to typical \xte\/ measurements shown in Fig.\ 
\ref{og_agn}. For example, NGC 5548 has the flux in the range of $\sim 5$--10, 
exposures of $\sim 20$--40 ks (Chiang et al.\ 2000; see also Section 
\ref{samples} below), IC 4329A has the flux of $\sim 15$, exposure of $\sim 70$ 
ks (Done et al.\ 2000), and 3C 120 has the flux of $\sim 6$, exposure of $\sim 
60$ ks (Eracleous et al.\ 2000). The results are shown in Fig.\ \ref{contours}b. 
We clearly see that the obtained distribution is very different from those in 
Fig.\ \ref{og_agn} and \ref{contours}a. It is also different from the contour of 
VE01 due to the better statistics.

In Fig.\ \ref{contours}b, we also show two representative error contours. We see 
that although they correspond to only $1\sigma$, they do provide a good measure 
of the spread of almost all the simulated points. This overestimate of 
statistical errrors by {\sc xspec} is due to the corresponding overestimate of 
background errors (added in quadrature to the total ones) discussed by N00 and 
VE01. In order to account for this effect, N00 did not include background errors 
in their data at all, and VE01 performed simulations to estimate the actual 
errors. In the case of the \ginga\/ data of ZLS99, the standard {\sc xspec} 
procedure was followed. Thus, the error contours on Fig.\ \ref{contours}a are 
somewhat overestimated. However, this is a conservative approach, which can only 
{\it reduce\/} the measured strength of an actual correlation but will not lead 
to an appearance of a spurious one. Therefore, we find the suggestion of Edelson 
\& Vaughan (2000) (based on the results of VE01) that the correlation of ZLS99 
could be spurious due to inadequate modeling of errors to be incorrect.

Results of simulations with a larger range of the input $\Gamma$ (and other 
assumptions similar to those above except for the systematic error being now 
0.01) are shown in Fig.\ \ref{contours}c. These results also show how the fitted 
values of $\Gamma$ and $\Omega/2\upi$ correlate with the ones assumed. We see no 
systematic effects here apart from the dispersion of the fitted value of 
$\Omega/2\upi$ increasing with the increasing $\Gamma$ (due to decreasing photon 
statistics at high energies).

To further study possible contributions from statistical effects, we have also
performed analogous simulations (using the exposure time, 5.4 ks, and response
as for a \ginga\/ observation of NGC 3516\footnote{The \ginga\/ spectra
available to us are already background-subtracted, and their errors have been
calculated including the uncertainty of the background. On the other hand,
simulations give errors neglecting the effect of the background. For this
particular source, this results in a decrease of the errors roughly
corresponding to an increase of the exposure by a factor of $\sim 2$. Thus, this
set of simulations corresponds to the actual exposure of $\sim 10$ ks, which is
still a half of the actual average exposure in the data of ZLS99 of 20 ks. On
the other hand, the effect of background errors in {\sc xspec} is generally overestimated (N00, VE01), and thus the present simulations may better
represent the actual uncertainties than those in Figs.\ \ref{contours}b-c.})
assuming $\Omega/2\upi$ to be random in the 0--1.5 range. Results for
100 simulated spectra are shown in Fig.\ \ref{contours}d. We see that the
obtained distribution is still very different from that of Fig.\
\ref{contours}a.

Our results confirm the finding of ZLS99 that the probability that a spectrum
with $(\Gamma,\,\Omega/2\upi)$ measured to be at the high end of the extent of
the correlation shown in Fig.\ \ref{contours}a would correspond in reality to a
point at a low end of the correlation is extremely low. ZLS99 noted that
$(\Gamma,\, \Omega/2\upi)$ contours are skewed and elongated (see Fig.\
\ref{contours}a), and thus used a statistical method taking into account the
correlated errors while looking for the actual functional dependences between
parameters. Based on comparison of the resulting values of $\chi^2$ (Bevington
\& Robinson 1992), ZLS99 determined the probability of the correlation appearing
by chance as $\sim 10^{-10}$. The exact value may be somewhat different  because
the model with an allowed $\Omega(\Gamma)$ dependence yields $\cnu\sim 2$ (which
reflects the intrinsic spread of the data). Still, the probability of reducing
$\chi^2$ from 312 (assuming constant $\Omega$) to 115 (allowing for a
phenomenological $\Omega[\Gamma]$ dependence as a power law) by chance is in any
case $\ll 1$.

\subsection{Intrinsic spread and small data samples}
\label{intrinsic}

We see in Figs.\ \ref{og_bhb}, \ref{og_agn} and \ref{contours}a that apart from
the correlation between $\Omega$ and $\Gamma$, there is also a clear intrinsic
spread of the values of those parameters in the observed $\Gamma$-$\Omega$
space. This spread is caused by varying model assumptions used in the references
on which Figs.\ \ref{og_bhb}, \ref{og_agn} are based, differences in calibration
of different instruments, and, last but not least, fluctuations due to some
physical effects (e.g., orientation) superimposed on the overall
$\Omega(\Gamma)$ dependence. The last effect is rather common in astrophysics,
in which it is very rare that any two quantities are correlated without any
influence of other parameters of the system. In the case of BH binaries, we see
a full width of $\Delta (\Omega/2\upi) \sim 0.2$ or so for a given $\Gamma$,
whereas it is $\sim 0.5$ for Seyferts.

Thus, a finding of no $\Omega$-$\Gamma$ correlation in a small sample cannot be
taken as a proof of the lack of it, even if the statistical quality of data is
very good. Furthermore, a sample with poor statistics cannot be considered a
basis of a proof for either presence or absence of the correlation. For example,
the data points for GX 339--4 by Wilms et al.\ (1999) have so large error bars
and so limited range of $\Gamma$ and $\Omega$ that no statement about the
correlation can be made at all. In the case of that source, the actual presence
of a strong correlation (see Fig.\ \ref{og_bhb}) has later been established by
Revnivtsev et al.\ (2001) and Nowak et al.\ (2002). Then, P02b claimed the
$\Omega$-$\Gamma$ correlation to be relatively weak in their sample of Seyferts.
However, the error bars in their data are large enough for their data to permit
the actual presence of a strong correlation. The same holds for the similar data
set of Matt (2001), who raised the issue of the influence of statistical effects
on the $\Omega(\Gamma)$ correlation found by him in the \sax\/ data but
presented no calculations to test it.

\subsection{Systematic effects}
\label{systematic}

Various systematic effects might affect the best fit values of the $\Omega$ and
$\Gamma$ and can lead to biased estimates and to contradiction between results
obtained by different instruments and using different spectral models. In
certain circumstances these effects could potentially lead to appearance of
spurious corelations between parameters. These effects are considered below.

N00 noted that a background subtraction with a systematic relative error with a
specific power-law dependence on energy (reaching 0.03 at 20 keV) may produce a
spurious $\Omega$-$\Gamma$ correlation in their \xte\/ data for NGC 7469.
However, it is easy to see that this would be the case only if the dominant
variability pattern corresponded to a pivot point at low energies, $\la 10$ keV.
Apart from the specific case of NGC 7469, we show in Section \ref{index_flux}
that the typical pivot energy in Seyferts is $\gg 10$ keV, which, in turn, would
produce an $\Omega$-$\Gamma$ anticorrelation. Thus, this explanation cannot be
general. Also, in Section \ref{samples} below, we reanalyze the data of
N00 with an updated PCA background model and recover the same correlation.
Furthermore, we consider it extremely unlikely that background would be
undersubtracted in the same way in AGNs observed by \ginga, \sax, and \xte.

Then, Weaver et al.\ (1998), and later P02b and Malzac \& Petrucci (2002, 
hereafter MP02), noted that a correction needs to be made to account for the 
high-energy cutoff in the spectra. Namely, the lower the cutoff the fewer 
incident photons are available for reflection, and then the fitted value of 
$\Omega/2\upi$ increases. A related important effect not noticed by P02b is the 
downward curvature of the incident model spectrum (present in the e-folded power law but not necessarily in Comptonization models) being compensated by an 
increase in the reflection strength. These two effects help to explain the 
offset between the \ginga\/ and \sax\/ results at low values of reflection 
noticed by Lubi\'nski \& Zdziarski (2001), and seen in Fig.\ \ref{og_agn}. On 
the other hand, P02b notice that the resulting correction to $\Omega$ cannot by 
itself remove the presence of the correlation.

We also point out here the existence of an important correction to the X-ray index not taken into account by P02b. It is the difference between the asymptotic low-energy index, $\Gamma_{\rm f}$, of an e-folded power law,
\begin{equation}
{{\rm d}{\dot N}\over {\rm d}E} \propto E^{-\Gamma_{\rm f}} \exp \left(-{E\over
\ef}\right),
\end{equation}
(where $N$ is the photon number), and the actual X-ray index
between two energies, $E_1$ and $E_2$. The latter equals 
$\Gamma= \Gamma_{\rm f} + \Delta\Gamma$, where
\begin{equation}
\Delta\Gamma = {E_2-E_1\over \ln E_2/E_1} \ef^{-1}\simeq {9\over \ef},
\label{delta}
\end{equation}
and the second equality corresponds to $E_1=3$ keV (limiting the range with
possible dominance of a soft X-ray excess or effects of ionized absorption) and
$E_2=20$ keV, used by ZLS99. At $\ef=400$ keV assumed by them,
$\Delta\Gamma\simeq 0.02$, which is negligibly small. Thus, the
correlation presented in ZLS99 concerns the actual hard X-ray index.

On the other hand, this model fitted to \sax\/ data on Seyferts by P02b (see 
also Matt 2001) yield generally lower values of $\ef$. Then, the fitted values 
of $\Omega/2\upi$ increase somewhat with respect to those corresponding to fits 
with no cutoff. For example, for $\ef=160$ keV, the fitted value of $\Omega$ 
increases by $\sim 50$ per cent (P02b). However, as pointed out above, the value 
of the index needs to be corrected as well, with $\Delta\Gamma=0.06$. In fact, 
the two points P02b found to disagree with the correlation of ZLS99 both require 
substantial corrections in $\Gamma$. In particular, $\ef=67$ keV fitted by P02b 
for Mrk 509 yields a rather large $\Delta\Gamma\simeq 0.13$, increasing the 
3--20 keV index from $\Gamma_{\rm f} =1.58$ (P02b) to $\Gamma \simeq 1.71$. 
Taking into account both corrections, to $\Omega$ and $\Gamma$, makes the \sax\/ 
results (Matt 2001; P02b) quite compatible with those from \ginga\/ of ZLS99 
(see Fig.\ \ref{og_agn}).

\begin{figure}
\centerline{\psfig{file=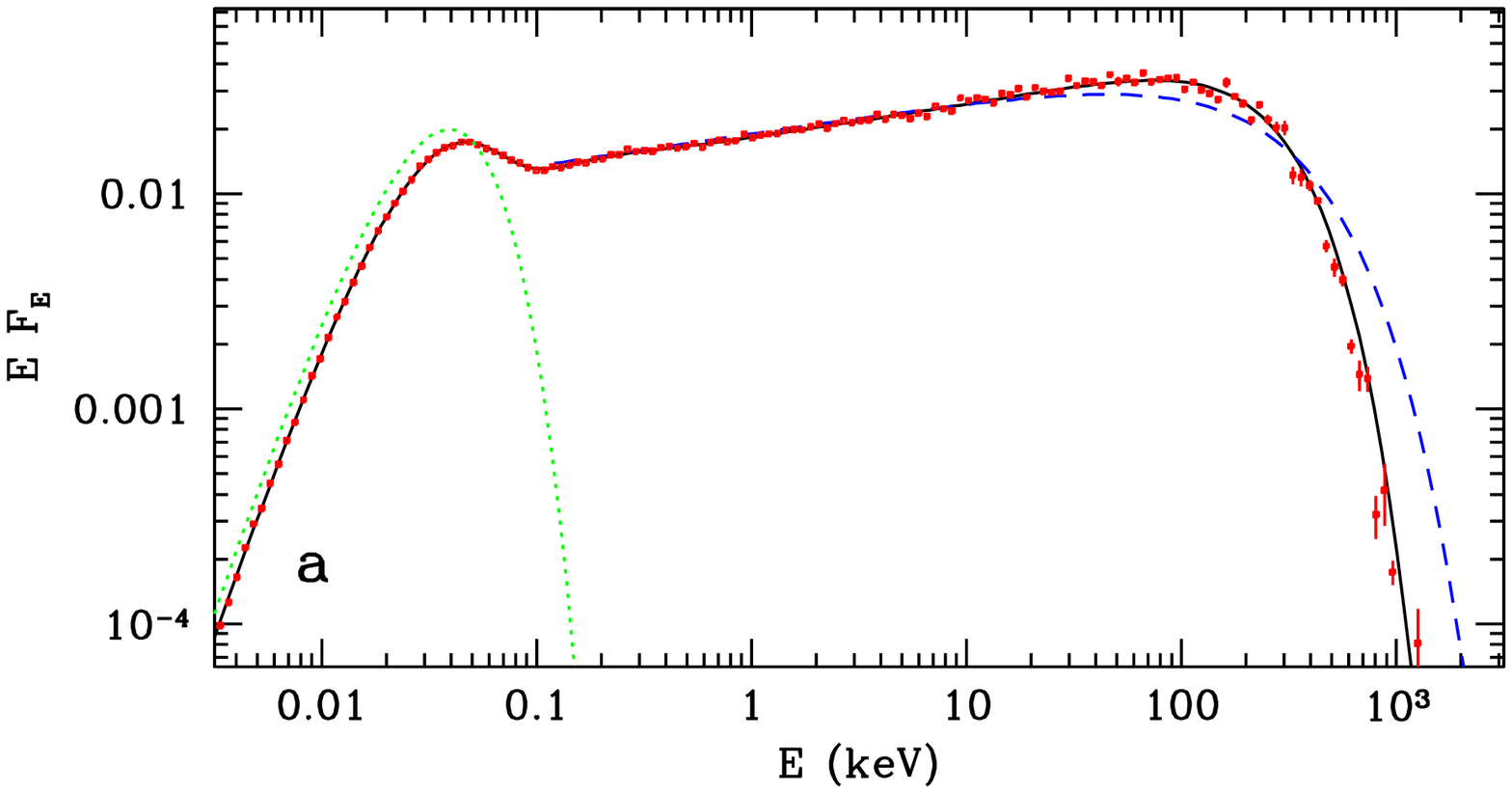,width=8.5cm}}
\centerline{\psfig{file=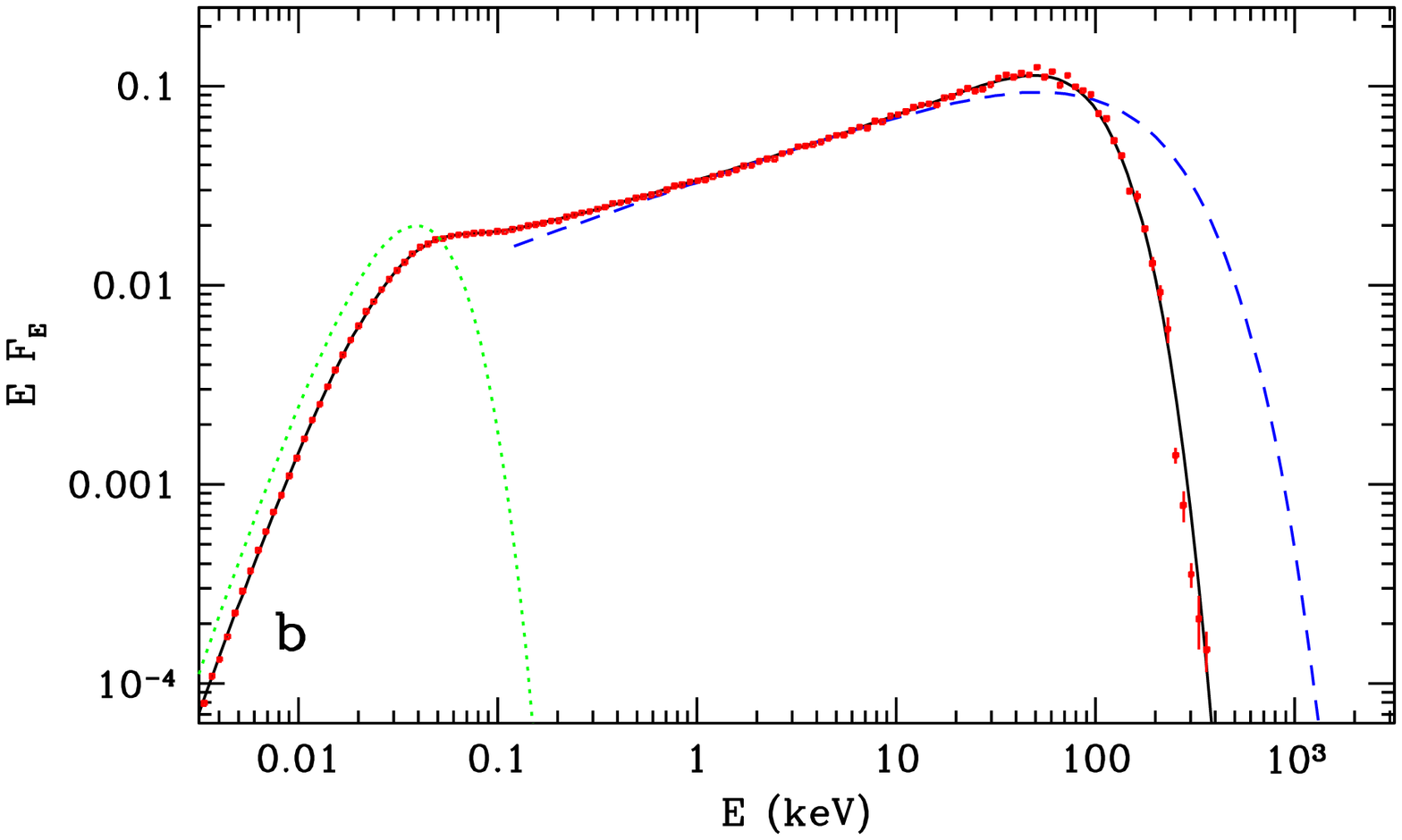,width=8.5cm}}
\caption{Spectra from isotropic thermal Comptonization compared to e-folded
power laws. Blackbody photons with $kT_{\rm bb}=10$ eV distributed uniformly in
a sphere are Comptonized by thermal electrons with the temperature, $kT$, and
the radial Thomson depth, $\tau$. The dotted, solid and dashed curves correspond
to the distribution of the input blackbody, Comptonized spectrum calculated by
the method of Poutanen \& Svensson (1996), and a characteristic e-folded power
law with the low-energy index of $\Gamma_{\rm f}$ and the folding energy of
$\ef$. The filled squares with vertical error bars correspond to Monte Carlo
results using the code of Gierli\'nski (2000). The parameters are: (a) $kT=100$
keV, $\tau=1$, $\Gamma_{\rm f}=1.85$, $\ef=300$ keV; (b) $kT=25$ keV, $\tau=4$,
$\Gamma_{\rm f}=1.65$, $\ef=150$ keV. We see that in neither case any e-folded
power law can approximate well the actual spectrum.
\label{Compton} }
\end{figure}

Furthermore, we point out that an e-folded power law is actually a very bad
model for thermal Comptonization, which physical model fits well the spectra of
Seyferts at high energies (e.g., Zdziarski, Poutanen \& Johnson 2000). For
spectral parameters characteristic to both Seyferts and BH binaries, thermal
Comptonization has a much sharper cutoff (following an extended, power-law like,
part of the spectrum) than that of the e-folded power law. This is illustrated
in Figs.\ \ref{Compton}a, b for two cases approximately spanning the range of
parameters of P02b. This further demonstrates that $\Gamma_{\rm f}$ of the
e-folded power law model has no direct physical meaning. The inadequacy of the
e-folded power law model was also noted by MP02, who have found that the sharper
cutoffs of the thermal Comptonization spectra lead to the increase of the value
of $\Omega$ being substantially less than that for the e-folded power law.

Concluding, results of fits with e-folded power laws need to be treated with
caution. We also note that Zdziarski et al.\ (1995) and Gondek et al.\ (1996)
obtained values of $\ef\ga 500$ keV for the average Seyfert spectra using data
from the \gro/OSSE detector (which affords coverage to significantly higher
energies than \sax), i.e., substantially higher than those from \sax.

In general, there are many other potential systematic effects that can affect
the absolute determination of $\Omega/2\upi$, e.g., due to an approximate
treatment of ionization of the reflecting medium. Then, in order to test the
reality of the correlation, we should ask whether a given set of fits properly
ranks the spectra by the reflection strength. This issue was addressed by
Gilfanov et al.\ (1999, see their fig.\ 7) and Revnivtsev et al.\ (2001, see
their fig.\ 3) in the case of Cyg X-1 and GX 339--4, respectively. They have
shown that the {\it ratio\/} of the count spectrum fitted with a higher value of
$\Omega/2\upi$ to that fitted with a lower value itself has the shape
characteristic to Compton reflection. This confirms that fits using relatively
simple reflection models do provide a correct ranking in the values of
$\Omega/2\upi$.

We would also like to comment on the use of a normalization of the reflection
spectrum of Nandra \& Pounds (1994) and  N00. They characterized the
strength of reflection by the 1-keV flux, $A_{\rm ref}$, of a power law with the
same $\Gamma$ as the incident one but normalized to $\Omega/2\upi=1$. Although
equivalent to using the measured $\Omega/2\upi$, this definition may lead
to a spurious systematic correlation between $A_{\rm ref}$ and $\Gamma$. To
illustrate it, let us consider $\Omega/2\upi=$ constant. A common variability
pattern of the X-ray power law spectra of both AGNs and BH binaries is variable
$\Gamma$ with an approximately constant pivot energy, $\ep$,
\begin{equation}
{{\rm d}{\dot N}\over {\rm d}E} =C \left(E\over E_{\rm p}\right)^{-\Gamma},
\label{spectrum}
\end{equation}
where $C$ is the normalization at $\ep$, and usually $E_{\rm p}\gg 1$ keV (see
Section \ref{index_flux} below). Then,
\begin{equation}
A_{\rm ref} \propto (E_{\rm p}/1\,{\rm keV})^\Gamma,
\end{equation}
i.e., $A_{\rm ref}$ is strongly positively correlated with $\Gamma$ even if the
actual reflection solid angle remains constant. Thus, we advise against using
this measure of reflection. (We note, however, that the specific data for NGC
7469 of N00 show $\Gamma$ uncorrelated with the X-ray flux, in which case the
above effect is not critical and probably does not affect substantially results
of that paper.)

\subsection{Single objects vs.\ samples}
\label{samples}

Another issue to bear in mind when considering evidence for and against the
$\Omega$-$\Gamma$ correlation are the distinctions between different classes of
sources and between samples of sources and repeated observations of a single
source. The evidence for the presence of this correlation in Seyferts was given
by ZLS99 mostly for broad-line Seyfert 1s as a class. Repeated observations of
single Seyferts give mixed results, sometimes showing the correlation in a given
source (e.g., Magdziarz et al.\ 1998 for NGC 5548; N00 for NGC 7469, also see
below), and sometimes not, e.g., in IC 4329A (Done et al.\ 2000) and some other
sources (P02b).

\begin{figure}
\centerline{\psfig{file=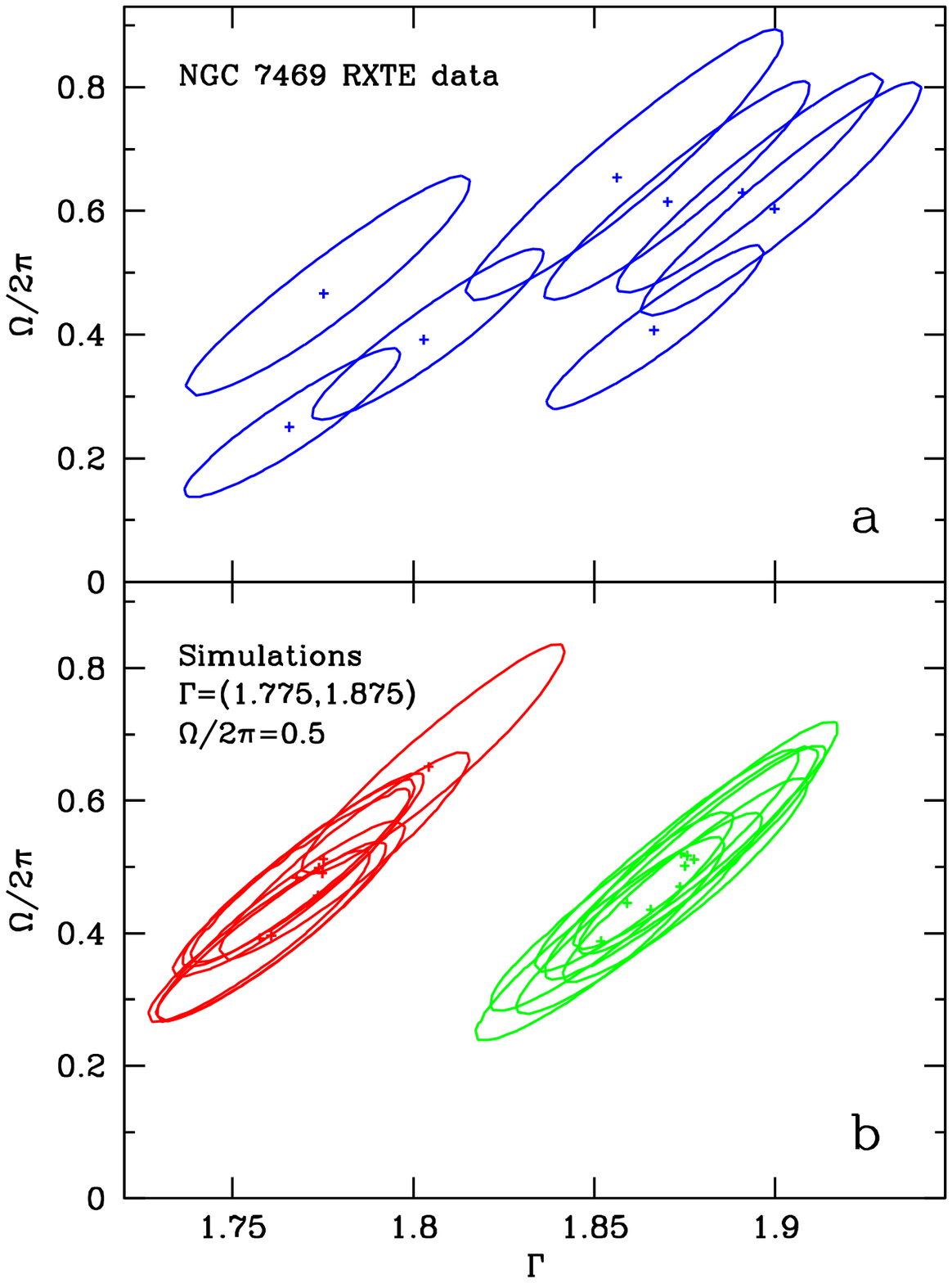,width=8cm}}
\centerline{\psfig{file=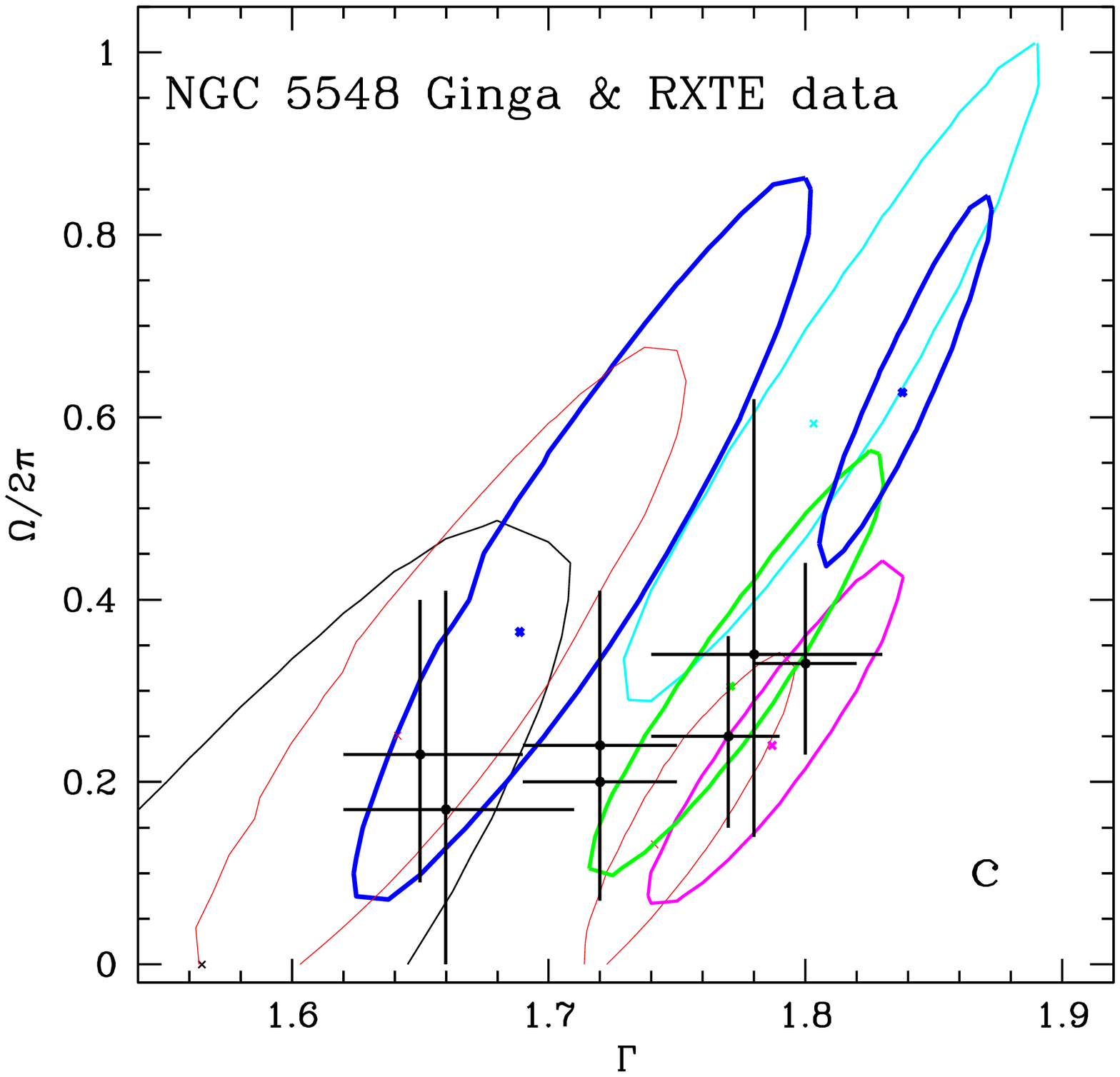,width=8cm}}
\caption{$\Omega$ vs.\ $\Gamma$ in two Seyferts. (a) The 1-$\sigma$ contours for
NGC 7469 obtained by us for the \xte\/ data (N00). (b) Simulations for NGC 7469.
We confirm the conclusion of N00 that the extent of the correlation cannot be
explained by statistical effects. (c) Contours for NGC 5548 from \ginga\/
(ZLS99), and error bars from \xte\/ (Chiang \& Blaes 2003).
\label{sy_ind} }
\end{figure}

Fig.\ \ref{sy_ind}a shows $\Omega$-$\Gamma$ contours for \xte\/ observations the 
Seyfert 1 NGC 7469 (the inclination was assumed to be $i=30\degr$). The data are 
the same as those of N00, but they have been reextracted by us using the 
LHEASOFT 5.2 version of the PCA response matrix and the background model. We 
used PCUs 0--4 whenever available and grouped the data in 4-day segments, and 
included a 0.5 per cent systematic error. This yields the exposure times of 
$\sim 50$--70 and $\sim 30$--50 ks for the PCUs 0--2 and 3--4, respectively. The 
contours clearly show a correlation (which would become even stronger after the 
reduction of the statistical errors discussed by N00 and VE01, see also Section 
\ref{statistical}). In order to test its reality, we have generated $>100$ 
simulated spectra (using the same exposures and fluxes as for the actual data), 
a selection of them shown in Fig.\ \ref{sy_ind}b for two assumed values of 
$\Gamma$. In all cases, we find the statistical $\Omega(\Gamma)$ dependence to 
be much steeper than the observed one. Thus, we confirm the corresponding 
conclusion of N00, who also ruled out the origin of the correlation in NGC 7469 
from statistical effects (see their appendix).

Fig.\ \ref{sy_ind}c shows the results for NGC 5548 from \ginga\/ by ZLS99
(contours) and from \xte\/ by Chiang \& Blaes (2003) (error bars). Note good
agreement between both sets of measurements after the update of the \xte\/
instrumental response by Chiang \& Blaes (2003) with respect to the original
result of Chiang et al.\ (2000), who claimed a disagreement with ZLS99.

In the case of BH binaries, the correlation is seen both in a number of
individual objects (Cyg X-1, GX 339--4, Nova Muscae in the hard state) as well
as in those sources considered together (Fig.\ \ref{og_bhb}). On the other hand,
it is certain that there are some BH binaries that do not obey the
$\Omega$-$\Gamma$ correlation. In particular, a disapearance of the correlation
with a decreasing Eddington ratio was pointed out in Section \ref{data}.

Although the data for Cyg X-1 in the soft state lie on the extrapolation of the
dependence for the hard state for this and other sources (Fig.\ \ref{og_bhb},
see also fig.\ 5 of Gilfanov et al.\ 1999), the existing data appear
insufficient to conclusively show the presence or absence of the
$\Omega$-$\Gamma$ correlation within the soft state. On the other hand, Rau \&
Greiner (2003) claimed the presence of an $\Omega$-$\Gamma$ correlation in
relatively soft states of GRS 1915+105. Then, Ballantyne, Iwasawa \& Fabian
(2001) did not find such a correlation in their sample of 5 narrow-line Seyfert
1s, which class of object is likely the extragalactic counterpart of the soft
state of BH binaries (Pounds, Done \& Osborne 1995). Similarly, Lamer et al.\
(2003) found no $\Omega$-$\Gamma$ correlation in the narrow-line Seyfert NGC
4051. The lack of the correlation in this object may be due to the domination of
the emission process by non-thermal electrons (see Section \ref{feedback}),
which possibility in soft-state sources was pointed out by ZLS99.

\section{X-ray index-flux correlations}
\label{index_flux}

\begin{figure}
\centerline{\psfig{file=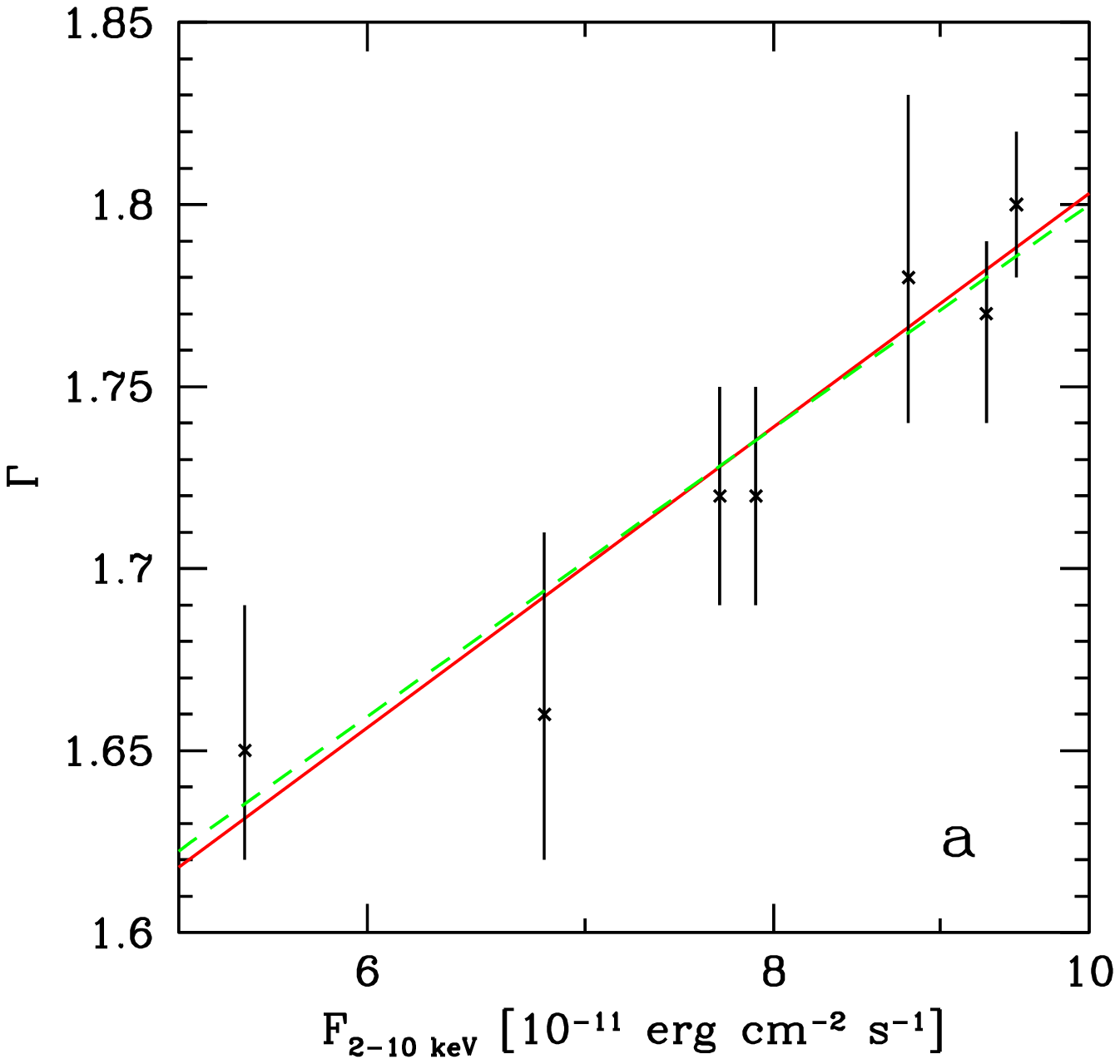,width=8.5cm}}
\centerline{\psfig{file=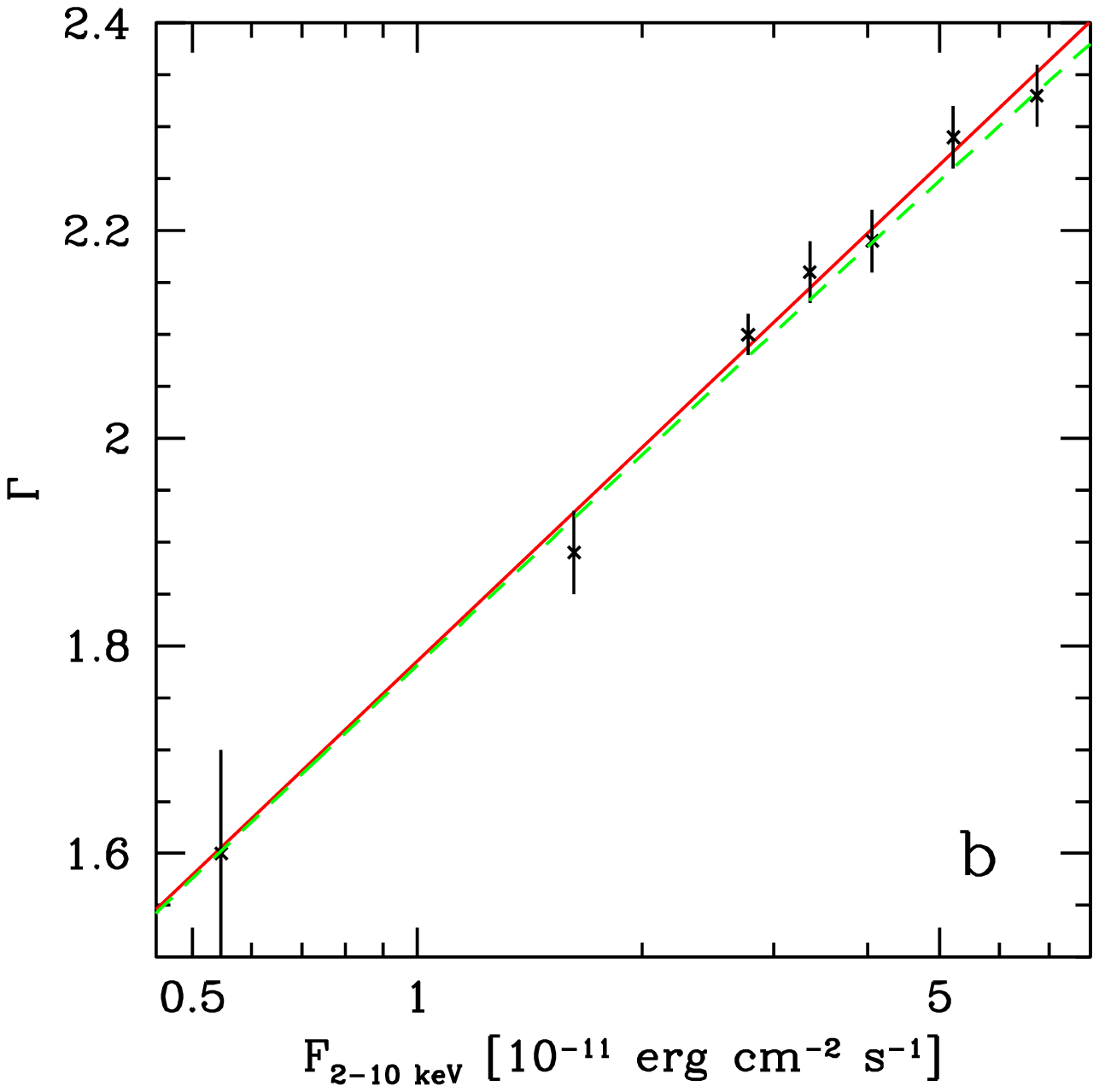,width=8cm}}
\caption{The $\Gamma$-$F$ correlation in \xte\/ observations of (a) NGC 5548,
and (b) NGC 4051. The data are from Chiang \& Blaes (2003) and Lamer et al.\
(2003), respectively. The solid lines show the best-fit linear dependence, and
the dashed lines show the dependence predicted by eq.\ (\ref{flux}) with
$\ep=180$ keV and 140 keV, respectively.
\label{fg} }
\end{figure}

Another common correlation concerning spectra of accreting BHs is that between
the X-ray spectral index, $\Gamma$, and the X-ray flux, $F$. A common situation
in the hard state of BH binaries and Seyfert 1s is $\Gamma$ showing an
increasing trend (although with often significant non-statistical scatter) with
$F$ in an X-ray energy range, $E_1$--$E_2$. Examples of this behaviour is shown
in Fig.\ \ref{fg} for the Seyferts NGC 5548 (Chiang \& Blaes 2003) and NGC 4051
(Lamer et al.\ 2003). The $\Gamma$-$F$ dependence can often be fitted by a power
law (e.g., Chiang et al.\ 2000; Done et al.\ 2000), also for the data shown in
Fig.\ \ref{fg}. The linear dependences of $\Gamma(\log F)$ in Fig.\ \ref{fg}
yield $\cnu<1$, i.e., the departures from the linear correlation are compatible
with being statistical only. On the other hand, the assumption of a constant
$\Gamma$ results in $\cnu=25/6$ and $138/6$, respectively, and the F-test
(Bevington \& Robinson 1992) yields the probability that the fit improvement is
by chance of $\sim 10^{-3}$ and $\sim 10^{-5}$, respectively. Thus, the
$\Gamma$-$F$ correlations are highly significant statistically, which conclusion
is also confirmed by application of the Spearman and Kendall rank correlation
tests. Similar power-law correlations are found, e.g., in the hard state of Cyg
X-1 on long time scales (Z02).

If the index increases from $\Gamma_{\rm h}$ to $\Gamma_{\rm s}$ while the
energy flux increases from $F_{\rm h}$ to $F_{\rm s}$, the two power laws
intersect at the pivot energy,
\begin{equation}
E_{\rm p}= E_1 \left[ { F_{\rm s}\over F_{\rm h}} \left( 2-\Gamma_{\rm
s}\over 2- \Gamma_{\rm h}\right) { (E_2/E_1)^{2-\Gamma_{\rm h}} -1
\over (E_2/E_1)^{2-\Gamma_{\rm s}} - 1 }
\right]^{1/(\Gamma_{\rm s}-\Gamma_{\rm h})},
\end{equation}
with the following substitution if either $\Gamma$ equals 2,
\begin{equation}
{(E_2/E_1)^{2-\Gamma} - 1\over 2-\Gamma}\rightarrow \ln {E_2\over
E_1}\,.
\end{equation}
If $F_{\rm s}/F_{\rm h}$ represents the ratio of the photon, rather than energy,
fluxes, then $2-\Gamma$ above should be replaced everywhere by $1-\Gamma$.
The resulting variable power law is given by equation (\ref{spectrum}). A
treatment of the effect of pivoting on average spectra and spectral variability
is given in Appendix A.

In the cases of the data for NGC 5548 and NGC 4051, $\ep\simeq 180$ keV and 140
keV, respectively. As shown in Appendix A, the assumption of a constant $\ep$
results in an almost linear dependence between $\Gamma$ and the logarithm of the
flux over some energy band. Indeed, the $E_1$--$E_2$ energy flux, given by
\begin{equation}
F_{E_1-E_2}= C\ep^{\Gamma}{E_2^{2-\Gamma}-E_1^{2-\Gamma}\over 2-\Gamma},
\label{flux}
\end{equation}
and plotted in dashed lines in Fig.\ \ref{fg}, almost coincide with the best-fit
linear dependences.

\begin{table*}
 \centering
 \begin{minipage}{130mm}
\caption{Pivoting in accreting black holes.}
\begin{tabular}{ccccccc}
\hline
Object & $E_1$--$E_2$ & $\Gamma_{\rm h}$ & $\Gamma_{\rm s}$ &
$F_{\rm s}/F_{\rm h}$\footnote{Energy flux ratio except when noted. } &
$E_{\rm p}$ & Reference \\
& [keV] &&&& [keV]\\
\hline
Cyg X-1 (hard state) & 3--12 & 1.4 & 2.0 & 3 & 40\footnote{Also observed in the
broad-band, 1.5--300 keV, variability (Z02). } & Z02\\
Cyg X-3 (hard state) & 3--100 & NA & NA & NA & $\sim 20$\footnote{Implied by the
3--100 keV variability. } & McCollough et al.\ (1999b)\\
3C 120 & 0.3--2 & 1.70 & 2.03 & 2\footnote{Count rate ratio assumed here to
represent photon flux ratio. } & 5 & ZG01\\
3C 390.3 & 1--10 & 1.70 & 1.90 & 1.8\footnote{Calculated using the 1-keV
normalization.} & 90 & Wo\'zniak et al.\ (1998)\\
3C 390.3 & 2--10 & 1.61 & 1.79 & 2.1 & 300 & Gliozzi et al.\ (2003)\\
IC 4329A & 2--10 & 1.90 & 2.07 & 1.9 & 200 & Done et al.\ (2000)\\
IC 4329A & 2--10 & 1.75 & 1.95 & 2.2 & 240 & Madejski et al.\ (2001)\\
MCG --6-30-15 & 2--10 & 1.8 & 2.2 & 3.3$^{d}$ & 80 & VE01\\
MCG --6-30-15 & 3--10 & 1.9 & 2.2 & 3.8\footnote{The (3--5)+(7--10) keV count
rate ratio assumed here to approximate the 3--10 keV
photon flux ratio. }  & 410 & P02a\\
Mrk 766 & 1--10 & 1.64 & 2.01 & 1.5$^e$ & 10 & Leighly et al.\ (1996)\\
NGC 3227 & 2--10 & 1.49 & 1.75 & 1.4 & 20 & Ptak et al.\ (1994)\\
NGC 3516 & 2--10 & 1.63 & 1.69 & 1.37 & 910 & Chiang (2002)\\
NGC 4051 & 3--10 & 1.3 & 2.6 & 26$^{f}$ & 60 & P02a\\
NGC 4051 & 2--10 & 1.60 & 2.33 & 12.3 & 140 & Lamer et al.\ (2003)\\
NGC 4151 & 2--10 & 1.4 & 1.7 & 6 & 2000 & Yaqoob et al.\ (1993)\\
NGC 5506 & 3--10 & 1.9 & 2.1 & 4$^{f}$ & 5000 & P02a\\
NGC 5548 & 2--10 & 1.65 & 1.80 & 1.73 & 180 & Chiang \& Blaes (2003)\\
NGC 5548 & 3--10 & 1.8 & 2.0 & 3.2$^{f}$ & 1700 & P02a\\
\hline
\end{tabular}
\end{minipage}
\label{t:pivot}
\end{table*}

Table 1 gives a number of examples of the pivot energy in Seyferts, as well as
in Cyg X-1 and Cyg X-3. Table 1 is based mostly on figures in published papers,
and thus the numbers there are approximate only. Still, they unambiguously show
that many Seyferts have the pivot energy at $\gg 10$ keV. A similar pattern is
shown by MCG --5-23-16 (Zdziarski, Johnson \& Magdziarz 1996). Although variable
X-ray absorption may affect the correlation for NGC 4151 found by Yaqoob et al.\
(1993), the weak variability found above 50 keV (Zdziarski et al.\ 1996; Johnson
et al.\ 1997) is consistent with the pivot at high energies.

\begin{figure}
\centerline{\psfig{file=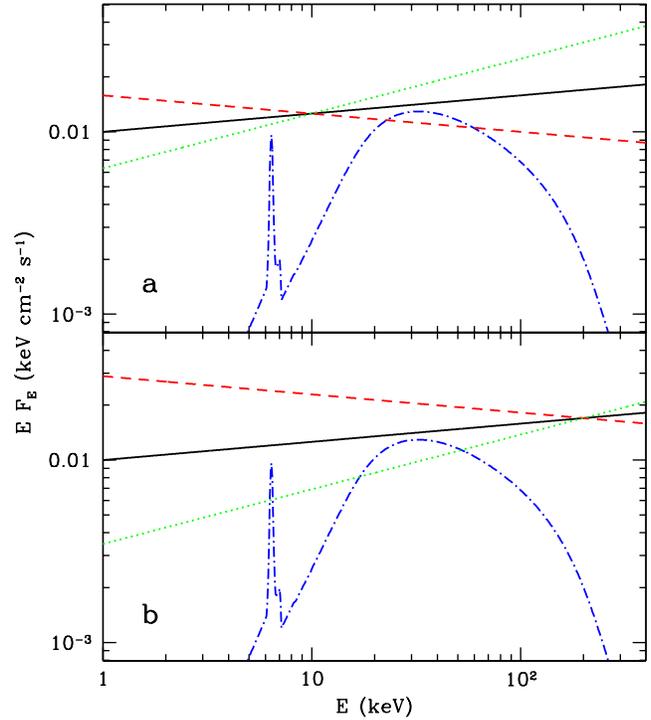,width=8.5cm}}
\caption{Schematic illustration of power law spectra with $\Gamma=1.7$, 1.9,
and 2.1 pivoting at (a) $E_{\rm p}=10$ keV and (b) $E_{\rm p}=200$ keV. The two
cases correspond to the total X-ray luminosity being approximately constant and
increasing with increasing $\Gamma$, respectively. The dot-dashed curve shows a
possible constant reflection component from a distant medium. If the total
spectrum is fitted by a power law and reflection, the relative reflection
fraction will increase and decrease with increasing $\Gamma$ in the cases (a)
and (b), respectively.
\label{pivot} }
\end{figure}

Table 1 gives two examples of AGNs with the pivot at $\leq 10$ keV, 3C 120 and
Mrk 766. In the cases of 3C 120 as well as Cyg X-1 in the hard state (where the
seed photon energies are much higher than those in AGNs, which results in the
pivot energy higher than that in 3C 120, see Z02), ZG01 and Z02 have shown that
the spectral variability is consistent with the bolometric luminosity of the
high-energy source being constant. This type of behaviour may also take place in
the Seyfert 1H 0419--577, where $\ep \sim 7$ keV (see fig.\ 3 in Page et al.\
2002). However, Table 1 shows that this type of behaviour is certainly not the
rule. The pivot at high energies implies the bolometric luminosity increasing
with the softening of the source, as illustrated in Fig.\ \ref{pivot}. This is
also confirmed by the 1--300 keV luminosity estimated by P02a for MCG --6-30-15,
NGC 5506 and NGC 5548 to increase by a factor of $\sim 3$ for $\Gamma$
increasing within the ranges given in Table 1. A similar increase is shown for
NGC 5548 in fig.\ 6a of Magdziarz et al.\ (1998). The results of Nowak et al.\
(2002, see their table 1 and fig.\ 3) imply the bolometric luminosity is
increasing with the increasing 3--9 keV flux in the hard state of the BH binary
GX 339--4. Some increase of the bolometric luminosity with increasing $\Gamma$
is also found in Cyg X-1 as the source gets close to a transition to the soft
state (Z02).

In the case of MCG --6-30-15, Shih, Iwasawa \& Fabian (2002) have found a 
$\Gamma$-$F$ correlation in \asca\/ data to appear to saturate above certain 
count rate. They have interpreted this behaviour as due to a superposition of 
two power law components with constant indices, and the normalization of the 
softer power law was allowed to vary. This model also allows to explain an 
approximate constancy of the Fe K$\alpha$ line flux with varying continuum flux 
in those data. Thus, this is an attractive model for MCG --6-30-15. On the other 
hand, data for many objects clearly show correlated hardness changes over rather 
broad bands (see the references above, e.g., P02a), which also rules out the 
above interpretation as general. Furthermore, the apparent saturation at a high 
$F$ seen by Shih et al.\ (2002) (and Merloni \& Fabian 2001, hereafter MF01) 
might be an artifact of plotting $\Gamma$ against the linear 
flux\footnote{Often, the ratio of instrument counts in two bands (hardness 
ration) is plotted as a function of a count rate. The advantages of this choice 
are independence of possible revisions of the instrument response (giving the 
conversion between counts and photons), which revisions are relatively common in 
X-ray astronomy, and no need to assume a spectral model. Another choice is to 
plot the fitted spectral index against the fitted energy flux. Although both 
quantities depend on the response, the advantage of this choice is an ease of 
comparison with physical models. On the other hand, VE01, MF01 and Shih et al.\ 
(2002) show hybrid plots with the fitted $\Gamma$ against the instrumental count 
rate. The last two papers show a comparison of the results with a physical model 
by MF01 assuming the X-ray count rate to be proportional to the energy flux. 
This is obviously not strictly correct for a variable X-ray slope, possibly 
leading to inaccuracies in comparing data with theory.} rather than $\log F$. As 
discussed above (see also Appendix A), variability with a constant pivot energy 
results in a linear dependence between $\Gamma$ and $\log F$, not $F$ itself. 
Indeed, the nonlinearity and saturation of $\Gamma$ in NGC 4051 claimed by Lamer 
et al.\ (2003) is specific to using a $\Gamma(F)$ plot, and it disappears 
completely in a $\Gamma(\log F)$ space (Fig.\ \ref{fg}b).

\begin{figure}
\centerline{\psfig{file=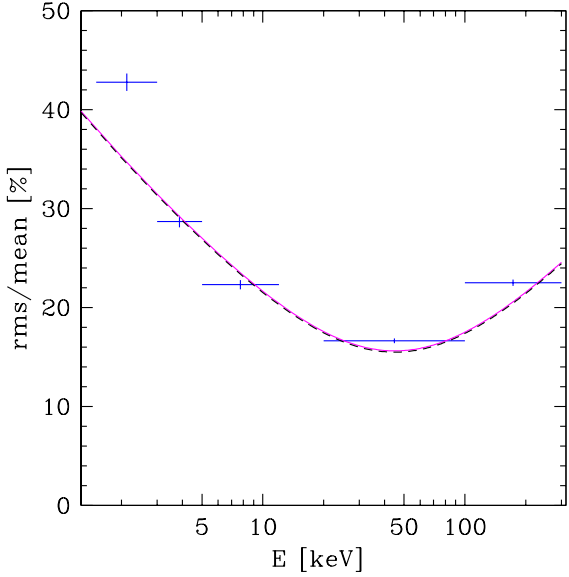,width=7.5cm}}
\caption{The rms variability in the hard state of Cyg X-1 in one-day averaged
measurements by the \xte/ASM and \gro/BATSE (Z02). The data (crosses) can be
modelled by either a power-law pivoting of equation (\ref{sigma_g}) with
$\ep=45$ keV, the standard deviation of $\Delta_\Gamma=
0.23$, and an additional $E$-independent variability with $\sigma_0=0.155$
(dashed curve), or a variable pivot energy (equation [\ref{sigma_f_2}]) with
$\ep^0=45$ keV, $\Delta_{\rm p}=0.71$, and $\Delta_\Gamma= 0.21$ (solid curve,
almost coincident with the dashed one). Note that the non-zero value of the
minimum rms of the latter model is entirely due to the dispersion of $\ep$, see
Appendix A.
\label{cyg_rms} }
\end{figure}

An important example of broad-band pivoting variability is given by $\sim 10^3$
of 1-day measurements of Cyg X-1 in the hard state over the 1.5--300 keV energy
range (Z02). Pivoting is seen here directly in the ASM and BATSE data showing a
linear $\Gamma(\log F)$ correlation at low energies and an anticorrelation at
high energies, not just being inferred from a $\Gamma$-$F$ plot. The pivoting
variability pattern can also be illustrated by energy-dependent fractional
variability. The solid and dashed curves in Fig.\ \ref{cyg_rms} show the
observed fractional rms variability as a function of energy is well modelled by
the theoretical rms due to a variable $\Gamma$ with either constant $\ep$ and
additional energy-independent variability or variable $\ep$. We also see that an
additional variable component is required at the lowest energies, which probably
can be identified with the observed soft X-ray excess (Ebisawa et al.\ 1996;
Frontera et al.\ 2001a; Di Salvo et al.\ 2001). Still, the 1.5--3 keV flux
remains strongly anticorrelated with the 100--300 keV flux (Z02).

Results for Seyfert 1s similar to that shown in Fig.\ \ref{cyg_rms} for Cyg X-1
have been obtained by Markowitz \& Edelson (2001). They show that the fractional
variability in the 2--4 keV band in all of 9 Seyfert 1s studied by them is
higher than that in the 7--10 keV band (for long-term measurements with 5-day
intervals). Given the unknown shape of the dependence of rms on $E$, it is not
possible to uniquely translate their results into the pivot energies. Still,
their results are compatible with $E_{\rm p} \ga 10$ keV.

The results above were obtained on time scales from days to years for both
Seyferts and BH binaries in the hard state. Interestingly, the variability
patterns in the two cases appear qualitatively similar. However, a month time
scale in Seyferts corresponds to $\sim 1$ s in BH binaries in units of the light
travel time accross the gravitational radius. Thus, the variability pattern
presented here for AGNs physically corresponds to that on short-time scales in
BH binaries. In agreement with this correspondence, Li, Feng \& Chen (1999) find
the 2--60 keV hardness anticorrelated with the flux (i.e., corresponding to a
positive $\Gamma$-$F$ correlation) in the hard state of Cyg X-1 at time scales
of $10^{-3}$ and 1--50 s. Thus, pivoting with a high pivot energy is likely to
take place in the BH binaries on short time scales as well. On the other hand,
the full picture of variability in the hard state is certainly more complicated
that this. For example, Li et al.\ (1999) and Wen, Cui \& Bradt (2001) find no
or weak $\Gamma$-$F$ correlation in Cyg X-1 on the 0.01--0.1 s and $\sim 1$-day
time scales, respectively, whereas it is very strong over hundreds of days
(Z02).

Then, spectral state transitions change the character of the correlation
qualitatively. In particular, the $\Gamma$-$F$ correlation becomes strongly
negative in the soft state of Cyg X-1 over time scales from 0.01 s to tens of
days (Li et al.\ 1999; Wen et al.\ 2001; Z02).

\section{Correlation between the X-ray and radio fluxes in BH binaries}
\label{radio}

A very interesting correlation between radio and X-ray fluxes in the hard state
of accreting BH binaries has been discovered recently (Brocksopp et al.\ 1999;
Corbel et al.\ 2000; Gallo et al.\ 2003; Markoff et al.\ 2003a). In particular,
Gallo et al.\ (2003) show a strong correlation between the 15 GHz flux and the
count rate from the ASM from daily measurements for 8 BH binaries in the hard
state. There are two possible origins of this correlation. One is that the level
of X-ray emission is related to the rate of ejection of radio-emitting clouds,
forming a compact jet (e.g., Mirabel et al.\ 1998; Corbel et al.\ 2000). Another
is that the X-ray emission of BH binaries is dominated by non-thermal emission
of the jet (Markoff, Falcke \& Fender 2001; Vadawale, Rao \& Chakrabarti 2001;
Markoff et al.\ 2003a, 2003b; Georganopoulos, Aharonian \& Kirk 2002).

There are many strong arguments against the second interpretation (see also
Poutanen \& Zdziarski 2003). The broad-band X\g\ spectra of BH binaries in the
hard state are very well modelled by thermal Comptonization and Compton
reflection (e.g., Gierli\'nski et al.\ 1997; Zdziarski et al.\ 1998; Frontera et
al.\ 2001a, b; Wardzi\'nski et al.\ 2002; Z02), see Fig.\ \ref{cygx1}. Apart
from those broad-band studies, the evidence for the presence of Compton
reflection from X-rays alone is extremely strong (e.g., Gilfanov et al.\ 1999;
CGR00; Revnivtsev et al.\ 2001; Nowak et al.\ 2002). This implies that the X-ray
emission is not strongly beamed away from the disc.

The thermal-Compton origin of the primary X-ray emission is strongly supported
by a remarkable uniformity of both the energy (within a factor of $\sim 2$) and
shape of the high-energy cutoffs of BH binaries in the hard state observed by
OSSE (Grove et al.\ 1998). This cutoff is naturally accounted for by
thermostatic properties of thermal Comptonization as well as \ee\ pair
production (e.g., Malzac, Beloborodov \& Poutanen 2001), as it corresponds to
the transition to relativistic temperatures. At higher temperatures, cooling
becomes extremely efficient and copious pair production starts. This reduces the
energy available per particle causing the temperature to decrease.

In the framework of the model with synchrotron emission of non-thermal electrons
accelerated in the first-order Fermi process in a shock,  the maximum energy of
the synchrotron photons is given by balance of the acceleration and energy loss
time scales. This gives the critical synchrotron energy corresponding to the
maximum Lorentz factor of the accelerated particles, $\gamma_{\rm max}$, as
\begin{equation}
E_{\rm c,max}={m_{\rm e} c^2\over \alpha_{\rm f} } {3^4 \sin \alpha\over 2^5}
{\beta_{\rm sh}^2\over \xi},
\label{emax}
\end{equation}
where $\beta_{\rm sh} c$ is the shock velocity, $\xi$ (satisfying $1/\beta_{\rm
sh}\ga \xi \ga 1$) gives the efficiency of the shock acceleration, $\alpha_{\rm
f}$ is the fine-structure constant, $\alpha$ is the pitch angle (with $\langle
\sin \alpha\rangle\sim 1/2$) and $m_{\rm e} c^2$ is the electron rest energy. We
see that for $\beta_{\rm sh}^2/\xi\sim 10^{-3}$ we can reproduce the high energy
cutoff of BH binaries  at $\sim 100$ keV, as pointed out by Markoff et al.\
(2001).

However, even with the fine-tuning of $\beta_{\rm sh}^2/\xi$, the
cutoff in the electron distribution is unlikely to be sharp. One effect is that
of electron energy loss, which makes the steady-state distributions to cut off
gradually (e.g., Kirk, Rieger \& Mastichiadis 1998). Also, the conditions in the
accelerating region (determining the efficiency parameter, $\xi$) are unlikely
to be completely uniform. Under many astrophysical circumstancies, the shape of
the tail of a distribution is given by an exponential cutoff (e.g., in a
Maxwellian). If we assume the power law electrons, $N(\gamma)=D\gamma^{-p}$, are
cut off as $\exp(-\gamma/\gamma_{\rm max})$, the resulting synchrotron spectrum,
$F_E=E {\rm d}\dot N/{\rm d}E$, is given by,
\begin{equation}
F_E= K\left(E\over E_0\right)^{1-p\over 2}\! \int_0^\infty\!\! {\rm d}y\,
y^{1+p\over 2} {\rm E}_{2+p}\left(\sqrt{\epsilon \over y} \right) {\rm K}_{5/3}
(y),
\label{syn_e}
\end{equation}
where $\epsilon =E/E_{\rm c,max}$, ${\rm E}_n$ is the exponential integral of
the order $n$, ${\rm K}_n$ is the modified Bessel function of the second kind,
the normalization constant is,
\begin{equation}
K={\sqrt{3} e^3 B D \sin \alpha \over h m_{\rm e} c^2},
\end{equation}
$B$ is the magnetic field strength, $e$ is the electron charge, $h$ is the
Planck constant, and $E_0$ is the constant factor in the
critical synchrotron energy, $E_{\rm c}$, corresponding to emission by electrons
with the Lorentz factor $\gamma$,
\begin{equation}
E_{\rm c}=E_0 \gamma^2,\quad E_0={3 e h B \sin \alpha\over 4\upi m_{\rm e} c}.
\end{equation}
The spectrum of equation (\ref{syn_e}) exhibits a rather gradual cutoff, in fact
{\it much\/} more gradual than that of a typical hard-state spectrum of Cyg X-1,
as as illustrated in Fig.\ \ref{cygx1}. Indeed, the synchrotron model of the
hard state of Cyg X-1 shown in fig.\ 3a in Markoff et al.\ (2003b) overestimates
the 1 MeV flux (McConnell et al.\ 2002) by a factor of 8 when matched to its 100
keV flux.

We have also studied various other forms of the electron high-energy cutoff. We
have found that anything more gradual than a sharp cutoff at $\gamma_{\rm max}$,
when
\begin{equation}
F_E= {2K \over p+1} \left(E\over E_0\right)^{1-p\over 2}
\int_\epsilon^\infty\!\! {\rm d}y \left(y^{1+p\over 2} -\epsilon^{1+p\over
2}\right) {\rm K}_{5/3} (y),
\end{equation}
gives photon spectra with cutoffs not strong enough to match the cutoff of Cyg
X-1, as illustrated in Fig.\ \ref{cygx1}.

A related issue is the spectral index of the spectrum. The model of Markoff et
al.\ (2001) relies on particle acceleration in shocks. As they note, the
accelerated electrons have then a power law index $>1.5$, where 1.5 corresponds
to highly relativistic shocks. As the terminal bulk motion of jets in BH
binaries is only mildly relativistic, with the typical bulk Lorentz factor of a
few, it is unlikely that a shock close to the base of a jet is highly
relativistic. Electron energy losses steepen the steady-state electron
distribution by unity (e.g., Kirk et al.\ 1998), to an index $p>2.5$, and the
corresponding photon index is $\Gamma=(p+1)/2$. Markoff et al.\ (2001) have been
able to fit this model to the spectrum of XTE J1118+480, which has $\Gamma\simeq
1.8$. However, many BH binaries have X-ray spectra harder than that (see Fig.\
\ref{og_bhb}). The resulting discrepancy is illustrated in Fig.\ \ref{cygx1},
where we assumed the limiting $p=2.5$ in the synchrotron model spectra shown.

In order to deal with this rather serious problem, Markoff et al.\ (2003a)
assumed that electron acceleration is continuous through the jet rather than
confined to a shock. Then, the steady-state electron index equals that of the
acceleration process (e.g., Kirk et al.\ 1998). Markoff et al.\ (2003a) do not
specify details of this acceleration process. A natural candidate seems to be
here the second-order Fermi acceleration in a turbulent medium (e.g., Jones
1994), in which case the electron index is constrained only by $p>1$. This
scenario is possible in principle, but then the acceleration depends on the
unknown properties of the accelerating medium and equation (\ref{emax}),
determining the high-energy cutoff, no longer applies.

The problems with the energy and shape of the high-energy cutoff are even more
severe for the non-thermal Compton model (Georganopoulos et al.\ 2002). It
requires that the product of the seed photon energy and the square of the
maximum electron energy is fine-tuned, the high-energy cutoff in the electron
distribution very sharp, and the seed photon distribution is very narrow. The
last condition is not fulfilled in any of the models of Georganopoulos et al.\
(2002), which rely on scattering of either disk or stellar blackbody photons.
Furthermore, the X-ray jet size in the model of Cyg X-1 of Georganopoulos et
al.\ (2002) is $10^{11}$ cm, which is clearly inconsistent with the peak of the
power spectrum per logarithm of frequency being at some occasions at $\sim 10$
Hz (e.g., Revnivtsev, Gilfanov \& Churazov 2000).

As discussed in Section \ref{ref_index}, the amplitude of Compton reflection and
the Fe K$\alpha$ flux (Section \ref{Fe}) imply that  dense and rather cold
material occupies a solid angle of $\Omega \sim\pi$ as viewed from the X-ray
source. Smearing of these components (e.g., Frontera et al.\ 2001a; Di Salvo et
al.\ 2001; GCR00) and their correlation with $\Gamma$ (Sections \ref{ref_index},
\ref{Fe}) clearly identify the reflector with the accretion disc (Section
\ref{theory}) and imply most of the X-ray source is within $\sim 30$--100
gravitational radii from the BH. On the other hand, models of Markoff et
al.\ (2001, 2003a, 2003b) and Georganopoulos et al.\ (2002) ignore that spectral
components.

Yet another piece of evidence against a substantial part of X-rays being
non-thermal is provided by spectral variability. In the case of Cyg X-1, the
ASM/BATSE data show a pivot around $\sim 40$--50 keV (Z02). This is consistent
with both the X-ray hardness-flux data and the rms variability strongly
increasing with decreasing X-ray energy, see Table 1 and Fig.\ \ref{cyg_rms},
respectively. The characteristic $\Delta_\Gamma$ from those data is $\sim
0.2$--0.3. This power-law spectral variability when extended to the turnover
energy (when the synchrotron source becomes optically thick) at $\sim 1$ eV
(e.g., Markoff et al.\ 2001, 2003a) would imply variability below this energy by
a factor of $\sim 10^2$. However, the range of the variability of the 15 GHz
flux correlated with the ASM flux in Cyg X-1 is only by a factor of several,
basically the same as the range of the variability of the ASM flux itself (Gallo
et al.\ 2003), contradicting the synchrotron origin of X-rays.

If both radio and X-rays were indeed due to non-thermal synchrotron emission
(Markoff et al.\ 2001, 2003a, 2003b), their observed variability pattern should
yield the X-ray rms virtually independent of energy. Then, in the model with
non-thermal Comptonization of photons from the companion star (Georganopoulos et
al.\ 2002), the seed photon flux is just constant, and the fractional
variability should increase with photon energy globally. These model predictions
are in strong disagreement with the data shown in Fig.\ \ref{cyg_rms}, and, in
particular, with the ASM 1.5--3 keV flux being stronly anticorrelated with the
100--300 keV flux from BATSE (Z02). A similar anticorrelation occurs in Cyg X-3
(McCollough et al.\ 1999b), and, in fact, the BATSE flux in that object is
anticorrelated with radio flux  (McCollough et al.\ 1999a).

On the other hand, if we extrapolate the rms dependence of Fig.\ \ref{cyg_rms}
to low energies and make a plausible assumption of the fractional rms being
$<1$, we obtain the characteristic energy of a fraction of keV. This is in very
good agreement with the temperature of seed photons for thermal Comptonization
observed in Cyg X-1 to be $kT_{\rm bb}\sim 0.15$ keV (Ebisawa et al.\ 1996; Di
Salvo et al.\ 2001).

All these results strongly support the interpretation of the correlated radio
emission as being due to ejection of clouds from the X-ray source, which could
be similar to coronal mass ejections observed at the Sun. This interpretation is
also supported by observed time lags of the radio emission with respect to the
X-rays, e.g., in GRS 1915+105 (Mirabel et al.\ 1998). In fact, the base of the
jet may be just the hot inner flow (e.g, Fender 2002). However, the electron
distribution in that base is still thermal, as argued above.

Finally, we note a problem with internal consistency in the X-ray jet model of
GRS 1915+105 of Vadawale et al.\ (2001). Namely, integrating the model
non-thermal synchrotron emission shown in their fig.\ 4 yields the total jet
luminosity of $\sim 10^{41}$ erg s$^{-1}$ (at a distance of 12 kpc,  jet
velocity of $0.9c$, and $i=70\degr$ adopted in that paper; note that this
emission is beamed away from the observer). On the other hand, the total jet
power given in that paper is $4.3\times 10^{39}$ erg s$^{-1}$. This power
includes the proton rest mass, and at the velocity of $0.9c$, the kinetic power
is $\sim 2\times 10^{39}$ erg s$^{-1}$. Thus, the jet radiative luminosity is
$\sim 50$ times the kinetic power, which strongly violates energy conservation.
Also, the radiative luminosity is usually much less than the kinetic power
(unless the jet stops entirely, in which case only the two luminosities can be
comparable), which furthermore increases the energy conservation problem.

\section{Theoretical Interpretation}
\label{theory}

\subsection{Feedback between cold and hot media}
\label{feedback}

The $\Gamma$-$F$ and $\Omega$-$\Gamma$ correlations occur likely due to
interaction between cold and hot media. The former correlation (also manifesting
itself as pivoting) is likely due to variability in the flux/luminosity of seed
soft photons (emitted by some cold medium) irradiating a hot plasma being much
stronger than the variability of the flux/luminosity from the hot plasma itself.
The main radiative process in the hot plasma needs to be thermal Compton
upscattering of the soft photons. Then, the larger the irradiating flux of seed
photons, the softer and stronger the X-ray spectrum. This variability pattern in
the case of a constant hot-plasma luminosity is shown, e.g., in fig.\ 3 of ZG01
and fig.\ 14 of Z02. The pivot point is then somewhere in the middle of the
broad-band spectrum. If the hot-plasma luminosity increases as well but slower
than the irradiating flux, the pivot is at high energies.

The emission of the cold medium irradiating the hot plasma may be partly (or
wholly) due to reprocessing of the emission of the hot plasma. Then, the
increased cooling of the hot plasma (which softens the X-ray spectrum) is
associated with more Compton reflection (accompanying reprocessing by a
Thomson-thick medium). This gives rise to an $\Omega$-$\Gamma$ correlation
(ZLS99; CGR00).

The two patterns can occur together or independently. If the variability of
soft seed photons is intrinsic (not from reprocessing), a $\Gamma$-$F$
correlation will not be accompanied by an $\Omega$-$\Gamma$ one. This may happen
in narrow-line Seyfert 1s. On the other hand, the variable seed photons may be
due to reprocessing but from emission of a highly variable hot plasma. Then,
there will be an $\Omega$-$\Gamma$ correlation but not a $\Gamma$-$F$ one, which
is the case for, e.g., Cyg X-1 on intermediate time scales (Gilfanov et al., in
preparation).

A number of specific geometries have been proposed. In one, there is a variable
radial overlap between the hot and cold accretion discs (ZLS99; Poutanen, Krolik
\& Ryde 1997). In another, the cold disc extends all the way to the minimum
stable orbit and the hot plasma forms a corona with a mildly relativistic
velocity directed either away from the disc or towards it (Beloborodov 1999a,
2001; Malzac et al.\ 2001). Also, static coronae have been considered in the
context of spectral correlations (e.g., Haardt, Maraschi \& Ghisellini 1997;
MF01).

\subsubsection{Variable overlap between hot and cold flows}
\label{flows}

ZLS99 have interpreted the $\Omega$-$\Gamma$ correlation as due to
feedback in an inner hot (thermal) accretion flow surrounded by an
overlapping cold disc. Then, the closer to the central BH the cold disc
extends, the more cooling of the hot plasma by blackbody photons (both
reprocessed and from intrinsic dissipation), and the softer the spectrum. The
effect of the cooling by the UV photons of the X-ray emitting plasma is seen,
e.g., in NGC 7469, where there is a positive correlation between the UV flux and
$\Gamma$ (N00). Also, the $\Gamma$-$F$ correlation (Section \ref{index_flux})
indicates the dominant effect of plasma cooling by a variable seed-photon flux.
At the same time, the cold disc subtends a larger solid angle from the point of
view of the hot plasma, and thus there are more reflection photons in the
spectrum. The solid curve in Fig.\ \ref{og_bhb} shows the prediction of a simple
version of this model (ZLS99).

This model also naturally accounts for the correlations of the reflection
strength with both the QPO frequency and the degree of relativistic smearing
seen in BH binaries (Gilfanov et al.\ 1999; GCR00; Revnivtsev et al.\
2001). The correlation with the QPO frequency is expected because that frequency
is very likely to be related in some way to the Keplerian frequency at the inner
edge of the cold disc, which increases with the decreasing disc inner radius. At
the same time, the closer the reflecting medium to the BH the higher the
degree of the relativistic smearing.

The model naturally explains the pivoting behaviour of the X-ray spectra
(manifesting itself in X-rays as a $\Gamma$-$F$ correlation) as driven by the
variable flux of irradiating seed photons (ZG01; Z02). Also, the specific
$\Gamma$-$F$ correlations observed in many accreting BHs indicate the bolometric
luminosity increases in those sources with the increasing $\Gamma$ (Section
\ref{index_flux}). This scenario can account for this behaviour if the inner
radius of the cold disk decreases with the increasing accretion rate. Such a
behaviour is indeed postulated in models of advection-dominated accretion (e.g.,
Esin et al.\ 1997) as well as predicted by models of accretion disk evaporation
(Meyer, Liu \& Meyer-Hofmeister 2000; R\'o\.za\'nska \& Czerny 2000). On the
other hand, it also appears likely that some instabilities can affect the disc
truncation radius even if the bolometric luminosity remains approximately
constant, which is observed in some objects (see Section \ref{index_flux}).

Chiang \& Blaes (2001, 2003) and Chiang (2002) have shown that detailed versions
of this model can also explain the overall optical/UV/X-ray variability in a few
Seyferts (NGC 3516, NGC 5548, NGC 7469). Note that their calculations require
the variable overlap to be mostly achieved by the radius of the hot plasma being
variable.

We note that this model does not readily explain $\Omega/2\upi>1$ sometimes
observed (see Figs.\ \ref{og_bhb}, \ref{og_agn}, \ref{contours}a). Furthermore,
scattering of the reflected photons in the hot flow will further reduce the
observed $\Omega$ (see a discussion in Beloborodov 2001). These problems can be
possibly resolved by anisotropy of the emission of the hot plasma or the outer
disc being concave. It is also possible that detections of such large reflection
are due to imperfection of the spectral models used and/or data inaccuracies.

In the framework of this model, radio emission arises from outflows in the hot
inner flow (e.g., Blandford \& Begelman 1999). Such a scenario is described,
e.g., in Fender (2002). Note that the jet emission in BH binaries is usually
quenched in the soft state (Brocksopp et al.\ 1999; Corbel et al.\ 2000), in
which the hot inner flow most likely disappears and is replaced by a hot corona
(e.g., Gierli\'nski et al.\ 1999).

\subsubsection{Static coronae}
\label{static}

Haardt et al.\ (1997) have studied spectral correlations in a static disc corona
model. They found that if the corona is dominated by \ee\ pairs, the 2--10 keV
spectral index, $\Gamma$ is rather insensitive to changes of the flux in the
same energy range, with typical $\Delta \Gamma\sim 0.2$ for a change of the flux
by 10. This is clearly much less than the observed spectral variability in many
sources (Section \ref{index_flux}), which rules out this model. The predictions
for coronae not dominated by pairs depend, in turn, on the choice of the coronal
optical depth, $\tau$. At $\tau\ga 0.3$, $\Gamma$ would decrease with the flux,
contrary to the data, but coronae with lower $\tau$ could be reconciled with
data.

Then, MF01 found that static patchy coronae could reproduce the observed
$\Gamma$-$F$ correlations if the luminosity of an active region increases with
its increasing size at a given height. The increased size increases the feedback
of soft radiation from the disc, which makes, in turn, the spectrum softer. MF01
and Shih et al.\ (2002) have shown that this model fits well the $\Gamma$-$F$
data for MCG --6-30-15 from \xte\/ and \asca, respectively.

On the other hand, this model yields an $\Omega$-$\Gamma$ anticorrelation rather
than correlation because an increased size at a given height (leading to the
softening of the spectrum, see above) increases the degree of obscuration of the
reflected radiation (Beloborodov 1999b, Malzac et al.\ 2001). This property is
also shared by static disc coronae in general, and it follows from the
dependencies shown by Haardt et al.\ (1997). Also, static coronae cannot by
themselves explain the formation of radio jets.

\subsubsection{Dynamic coronae}
\label{dynamic}

An interpretation of the $\Omega$-$\Gamma$ correlation alternative to the
variable overlap of two accretion flows (Section \ref{flows}) is that with
mildly relativistic coronal inflows/outflows (Beloborodov 1999a; Malzac et al.\
2001). The higher the speed of the coronal outflow, the less feedback with the
underlying disc and the harder the spectrum. Inflows can, in turn, account for
$\Omega/2\upi>1$.

It is not clear how to account for the $\Gamma$-$F$ correlation in this model.
This correlation requires that the observed luminosity of the hot plasma either
stays constant or increases with a decrease of the outflow velocity. On the
other hand, an opposite behaviour of constant luminosity of the seed photons and
variable luminosity of the hot plasma (which results in a $\Gamma$-$F$
anticorrelation) is seen in the soft state of Cyg X-1 (Churazov, Gilfanov \&
Revnivtsev 2001; Z02), in which case coronal models are widely accepted (e.g.,
Poutanen et al.\ 1997; Gierli\'nski et al.\ 1999; Churazov et al.\ 2001).

In the framework of this model, radio emission may arise from the jet being
formed by the coronal outflow (Merloni \& Fabian 2002). However, the
quenching of the radio emission in the soft state (Brocksopp et al.\ 1999;
Corbel et al.\ 2000) is not readily explained in this model.

\subsubsection{Further physical implications}
\label{implications}

\begin{figure}
\centerline{\psfig{file=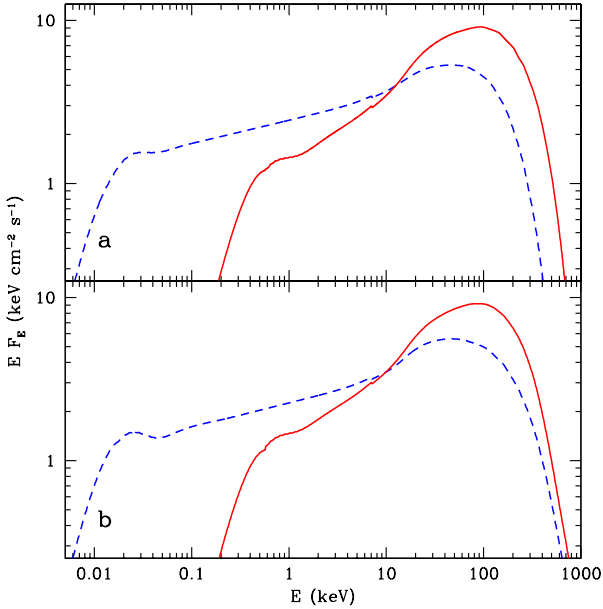,width=8.cm}}
\caption{Illustration of the effect of changing $kT_{\rm bb}$ of the
seed photons from 5 eV (dashed curves; characteristic to AGNs) to 150 eV
(solid curves, characteristic to BH binaries). In both cases the
Comptonization amplification factor, $A= 10$. The plasma is dominated by
(a) electrons and protons ($e^- p$) and (b) \ee\ pairs. The $e^- p$ plasma
parameters are $\tau=1.5$, $kT=57$ keV, and $\tau=1.5$, $kT=87$ keV, in the AGN
and BH binary cases, respectively, which yields the corresponding 2--10 keV
spectral indices of $\Gamma =1.85$ and 1.63. The \ee\ plasma parameters are
$\tau=1.0$, $kT=84$ keV, and $\tau=1.6$, $kT=72$ keV, in the AGN and BH binary
cases, respectively. In all cases, $\Omega/2\upi=0.4$ at 
$i=30\degr$.
\label{bb} }
\end{figure}

The feedback models explain the $\Omega$-$\Gamma$ correlation in terms of
Comptonization of blackbody photons by a hot thermal plasma. The effective
reflection solid angle in this model can be linked to the geometry of the
source, which will also determine the amplification factor, $A$, of the
Comptonization process. On the other hand, the spectral index follows from the
energy (and \ee\ pair) balance. Since the characteristic blackbody seed photon
energy is much lower in AGNs than in BH binaries, a generic prediction of these
models is that for a given $\Omega$ the value of $\Gamma$ will be higher in AGNs
than that in BH binaries (Beloborodov 1999b; Malzac et al.\ 2001). This is
illustrated in Fig.\ \ref{bb} (which calculations were performed using a code of
Coppi 1999). When the plasma compactness ($\propto L/R$, where $L$ and $R$ are
the luminosity and size, respectively), is low, \ee\ pair production is
negligible and the energy balance is satisfied by adjusting the electron
temperature, $kT$, see Fig.\ \ref{bb}a. On the other hand, at high compactness,
the energy balance is achieved by adjusting the Thomson optical depth of the
pairs, see Fig.\ \ref{bb}b. (In the calculations shown in Fig.\ \ref{bb}b, we
assumed the presence of a weak high-energy power-law tail, containing 0.05 of
the total \ee\ energy beyond the Maxwellian electron distribution, as suggested
by the results of McConnell et al.\ 2002.) The calculations shown in Fig.\
\ref{bb}, with $\Gamma$ changing from 1.63 to 1.85 at the assumed
$\Omega/2\upi=0.4$, explain well the difference between the $\Gamma$-$\Omega$
correlations for the BH binaries (Fig.\ \ref{og_bhb}) and AGNs (Fig.\
\ref{og_agn}).

An important issue here is also the possible presence of a correlation between 
$\Gamma$ and $kT$ for a given $kT_{\rm bb}$. The feedback models require that 
the parameters of the hot thermal plasma adjust to variable cooling to satisfy 
the energy balance. If $\tau$ is constant, this can be done only by changing 
$kT$. Then, an increased cooling yielding an increase of $\Gamma$ is associated 
with a decrease of $kT$. On the other hand, it is possible that $\tau$ also 
changes, e.g., responding to changing accretion rate, which would produce a more 
complicated behaviour (e.g., Chiang \& Blaes 2001, 2003). Also, if \ee\ pairs 
are present, the main adjusting parameter is $\tau$. In fact, in the example in 
Fig.\ \ref{bb}b, the softening of the spectrum is associated with an increase 
(rather than decrease) of $kT$.

So far there are only limited observational data on correlation between $\Gamma$
and the high-energy cutoff. P02b found a positive correlation between $\Gamma$
and the e-folding energy in their sample of Seyfert 1s, which would be contrary
to expectations of the simplest model with constant $\tau$ and the variability
of $\Gamma$ due to changing $kT$. However, MP02 found that this correlation may
be an artifact of the e-folded power-law model when used to fit thermal-Compton
spectra. They simulated actual thermal Comptonization spectra at a {\it
constant\/} $kT=100$ keV and a range of $\tau$, and then found that the
simulated spectra when fitted by an e-folded power law yield a correlation
resembling that of P02b.

\subsection{Reflection from strongly ionized disc}
\label{ion}

Done \& Nayakshin (2001) have shown that the optical depth of a highly ionized
surface layer on top of a strongly irradiated accretion disc increases with the
increasing hardness of the irradiating spectrum. This gives rise to an apparent
$\Omega$-$\Gamma$ correlation as there is less unscattered reflection for harder
spectra (although the actual reflection solid angle is $\Omega=2\upi=$
constant). A separate mechanism is then needed to account for the $\Gamma$-$F$
correlation, similarly to the case of the coronal outflow model.

\begin{figure}
\centerline{\psfig{file=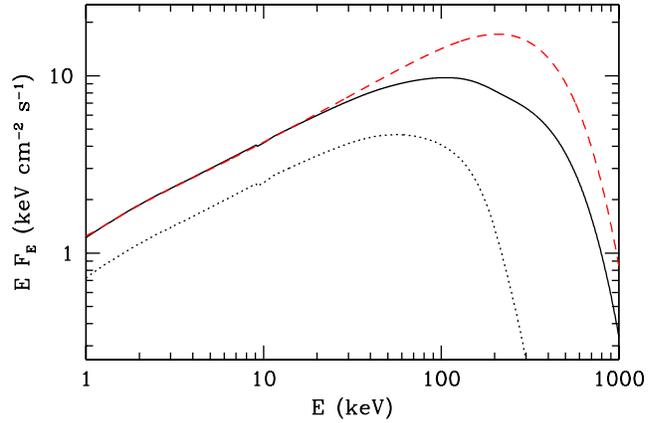,width=8.5cm}}
\caption{Comptonization spectrum with negligible reflection (dashed curve)
compared to a spectrum with $\Omega/2\upi =1$ and the reflecting medium being
almost fully ionized (solid curve). The dotted curve shows the ionized
reflection component alone. We see that the form of the high-energy cutoff in
the reflection spectral component leads to the total spectrum being distinctly
different from the Comptonization spectrum alone (although the two coincide at
low energies). See Section \ref{ion} for details.
\label{ionref} }
\end{figure}

A diagnostic that could yield the actual solid angle of the reflector regardless
of the ionization level is a measurement of reflection at high energies. Namely,
the reflection spectrum is cut off at high energies with a functional form
independent of the ionization state (White, Lightman \& Zdziarski 1988). Fig.\
\ref{ionref} illustrates this point for an incident spectrum from Comptonization
(Poutanen \& Svensson 1996) with $kT=120$ keV and $\tau=2$ in spherical
geometry, yielding the 2--10 keV index of $\Gamma=1.5$. This $\Gamma$
corresponds to the case with the lowest fitted neutral reflection from a
strongly ionized medium with the actual $\Omega/2\upi=1$ in Done \& Nayakshin
(2001). To illustrate this effect, we assumed that the reflecting medium is so
strongly ionized that virtually no Fe K edge appears in the reflection component
(dotted curve in Fig.\ \ref{ionref}). Although the reflected component is simply
a power law at low energies, it still does have a high energy cutoff due to
Klein-Nishina effects. The solid curve shows the total spectrum. Then, the
dashed curve shows the incident spectrum without reflection normalized to
coincide with the spectrum from the strongly ionized medium. We see that
although the spectra at $\la 20$ keV are indeed barely distinguishable, the form
of their high-energy cutoffs is very different. Thus, this model can be tested
using broad-band X\g\ spectra extending to several hundred of keV.

A set of $\sim 1$--1000 keV spectra of Cyg X-1 in the hard state was analyzed by
Gierli\'nski et al.\ (1997), who found the form of the high-energy cutoff
compatible with thermal Comptonization. Although those authors have not consider
the ionized-reflection model, those spectra have not shown any hint of the
softening of the cutoff of the type seen in the solid curve in Fig.\
\ref{ionref}; rather they indicated some additional hardening (interpreted by
them as due to the presence of an additional spectral component from saturated
Comptonization).

Independently of the present work, Barrio, Done \& Nayakshin (2003) have
recently fitted the ionized reflection model to broad-band spectral data from
the PCA and HEXTE detectors of \xte\/ for Cyg X-1. They have found this model to
be ruled out by the lack of a break in the HEXTE data, in favour of the model
with a truncated disk and intrinsically weak reflection. Future tests of this
model are highly desirable, e.g., using data from {\it INTEGRAL\/} and {\it
ASTRO-E2}.

The effect shown in Fig.\ \ref{ionref} also demonstrates that the albedo, $a$,
of a fully ionized medium energy-integrated over a spectrum characteristic to
accreting BHs is still substantially less than unity (see also White et al.\
1988). Indeed, in the specific case in Fig.\ \ref{ionref}, $a=0.48$, only
moderately larger than the corresponding albedo for a neutral reflector,
$a=0.34$. This shows that a fully ionized surface layer of an accretion disc
would radiate substantial soft thermal flux even in the absence of an internal
dissipation. This may rule out the ionized-disc model of the hard state of Cyg
X-1 of Young et al.\ (2001). In general, the resulting constraints on the
geometry will be relatively similar to those in the case of a nearly-neutral
disc (e.g., Stern et al.\ 1995), and, in particular, will still rule out the
presence of a homogeneous corona above the disc.

\subsection{Time lags and reflection}
\label{lags}

N00 have pointed out that the presence of a constant reflection component from a
distant medium (presumably a molecular torus) in AGNs together with a variable
power law (from hot plasmas in the vicinity of the BH) can lead to an
apparent correlation between $\Gamma$ and $\Omega$ if reflection in the fitted
model is tied to the variable power law. In particular, when the power law
varies with a characteristic pivot energy of $\la 10$ keV, the fitted
$\Omega$-$\Gamma$ dependence will be positive (i.e., the softer the spectrum,
the larger the relative contribution of reflection), as illustrated in Fig.\
\ref{pivot}a.

This idea has been developed in detail by MP02. They have considered cases with
the pivot energy of 2, 5 and 10 keV, and have shown that the $\Omega$-$\Gamma$
dependence in Seyferts (Fig.\ \ref{og_agn}) can be qualitatively explained by
this type of model. However, we have shown in Section \ref{index_flux} that the
pivot energy in many Seyferts is $\ga 100$ keV. Then, if this effect were
important for a large fraction of Seyferts, an $\Omega$-$\Gamma$ {\it
anticorrelation\/} (see Fig.\ \ref{pivot}b) would be observed for a large
fraction of sources, in conflict with the data. This rules out the origin of the
global $\Omega$-$\Gamma$ dependence in Seyferts from this effect. On the other
hand, this effect may be important in some Seyferts, and may explain
$\Omega$-$\Gamma$ anticorrelation in repeated observations of certain objects
(Done et al.\ 2000; P02a).

As noted by MP02, this effect cannot play any role in BH binaries. Thus, this
model requires completely different origin for the $\Omega$-$\Gamma$
correlations in those objects and in AGNs.

\section{The F{\lowercase{e}} K$\bmath{\alpha}$ emission}
\label{Fe}

The results presented in Section \ref{ref_index} were mostly obtained by fits to
the Compton-reflection continuum. However, the process of bound-free absorption
in the reflecting medium is usually followed by emission of a fluorescent Fe
K$\alpha$ line, which equivalent width, $\wfe$, can be tied to that of the
reflection (George \& Fabian 1991; \.Zycki \& Czerny 1994). Conversely, if
relativistic broadening is seen in the line emitted by the reflecting medium,
the same broadening should be seen in the reflection spectrum. Thus, the
measured parameters of the Fe K$\alpha$ line should be in agreement with those
of reflection. Such an agreement appears to be seen in BH binaries (Gilfanov et
al.\ 1999; GCR00; Revnivtsev et al.\ 2001). Some disagreement in GX 339--4
claimed by Nowak et al.\ (2002) appears to be largely caused by a diffuse line
component from the interstellar medium dominating in X-ray weak states of this
source (Wardzi\'nski et al.\ 2002).

On the other hand, the situation is more complex in Seyferts. Unlike BH
binaries, Seyferts possess significant Fe K$\alpha$ line component from distant
matter, e.g., the broad-line regions or molecular torii. Often, that distant
matter is Thomson thin, and then it gives rise to the line but not to an Fe K
edge, see, e.g., section 7.1 and fig.\ 13 in Wo\'zniak et al.\ (1998). As shown
in that figure, the optically-thin reflection component is then relatively
negligible. Then, the presence of two line components, one constant (with or
without associated reflection) and one variable in addition to a variable
intrinsic component may cause a rather complex behaviour of variability. For
example, the presence of constant line and reflection and a power law pivoting
at $\sim 10$ keV leads to an anticorrelation of $\wfe$ and $\Omega/2\upi$
measured with respect to the power law (see Fig.\ \ref{pivot}a, also MP02). Such
an effect might have been seen by Chiang et al.\ (2000) in NGC 5548.

Given this complexity, it is important to self-consistently account for various
components of the line and reflection, their relativistic broadening, and their
connection to the incident spectral component. A self-consistent treatment
should in general include at least two line components, narrow and broad, with
the broadening (e.g., Gaussian or disc-like) and redshift of the latter applied
in the same way to the reflection (e.g, \.Zycki et al.\ 1998, 1999; GCR00,
Lubi\'nski \& Zdziarski 2001; Zdziarski et al.\ 2002a). The strength of
reflection should then be tied to that of the line (George \& Fabian 1991;
\.Zycki \& Czerny 1994). The dramatic effect of inconsistent treatment of the
line and reflection is shown for the case of the disc broadening (Fabian et al.\
1989) in fig.\ 3 of Zdziarski et al.\ (2002a).

Then, Shih et al.\ (2002) showed that Fe K$\alpha$ line flux in MCG --6-30-15
remained approximately constant with varying 3--10 keV \asca\/ count rate.
However, the formation of the line is due to photons only above 7.1 keV or more.
In the case of MCG --6-30-15, $\Gamma\sim 2$, and then the fraction of the
measured \asca\/ counts above 7 keV in their data is only $\sim 0.03$. The
fraction of counts above 8 keV, appropriate for a moderately ionized disc,
possibly present in MCG --6-30-15 (e.g., Ballantyne \& Fabian 2001), is as low
as $\sim 0.01$. Thus, the lack of a correlation between the 3--10 keV counts and
the line photon flux cannot be taken as conclusive evidence against the Fe
K$\alpha$ photons being due to irradiation by photons from the observed
continuum. Similarly, the fraction of counts above 7 and 8 keV in the MCG
--6-30-15 data of VE01 is only $\sim 0.2$ and $\sim 0.1$, respectively. The data
of VE01 imply a pivot energy at about 80 keV (Table 1); thus, the variability of
the flux above $\sim 7$--8 keV will be in general less than that in the 2--10
keV band used by them (as also found by Markowitz \& Edelson 2001 for Seyfert 1s
in general).

As noted in Section \ref{index_flux} (as well as by MP02), there are many AGNs 
with the pivot at $\la 10$ keV (e.g., 3C 120). Then, the 7--10 keV photon flux 
may be anticorrelated with the 2--10 keV one. Even if the pivot energy is in 
general at $\gg 10$ keV, occasionally the same object may show a pivot at $\la 
10$ keV, as in the case of NGC 5548 (Nicastro et al.\ 2000). Thus, it is 
important to use the band above the Fe K edge in studies of the continuum-line 
correlations.

Self-consistent treatments of the line and reflection was applied, e.g., to the
Seyfert galaxy IC 4329A by Done et al.\ (2000), who found no disagreement
between the line and reflection. A similar method was applied to NGC 4151 by
Zdziarski et al.\ (2002a), who also found the broad-line component compatible
with reflection, in spite of previous claims to the contrary (e.g., Wang, Zhou
\& Wang 1999). Then, Lubi\'nski \& Zdziarski (2001) have applied the same method
to average \asca\/ spectra of Seyfert 1s and found the dependence of the $\wfe$
of the broad component of Fe K$\alpha$ lines to obey the dependence on $\Gamma$
compatible with that of the reflection continuum. Although Yaqoob et al.\ (2002)
have pointed out that the statistical weights used by Lubi\'nski \& Zdziarski
(2001) were not justified statistically, the effect of their correction appears
to be only minor (Lubi\'nski, in preparation). Indeed, the results of P02b for
Seyfert 1s observed by \sax\/ show an increase of $\wfe$ with $\Gamma$ very
similar to that obtained by Lubi\'nski \& Zdziarski (2001).

\section{Conclusions}

We have considered correlations between various spectral properties of accreting
BHs in Seyfert galaxies and X-ray binaries, with particular emphasis on the
correlations between the X-ray spectral index, strength of Compton reflection
and the X-ray flux. The main results can be summarized as follows.

Using published data of the observations of Seyferts with \ginga, \xte\/ and
\sax, we have critically re-evaluated the evidence for presence of correlation
between the X-ray spectral index and strength of Compton reflection. We
conclude that when considering a large number of observations of a large sample
of objects, the existence of {\it global\/} correlation between these two
parameters is established beyond any reasonable doubts.

Smallness of the error bars in comparison with the extent of the correlation and
good agreement of the results obtained by various satellites confirm that the
correlation cannot be an artifact caused by statistical or systematic effects.
The ratios of the spectra with different values of reflection in BH binaries
demonstrate that the correlation cannot be a consequence of a trivially
inadequate spectral model. We note, however, that the particular values of the
spectral index and, especially, of the strength of the Compton reflection, do
depend on the details of the spectral approximation. This fact should taken into
consideration when comparing results obtained by different authors.

The $\Omega$-$\Gamma$ correlation shows significant spread, larger than the
statistical uncertainties of the data. It is not clear at present to which
degree this spread is intrinsic to the sources and to which degree it is due to
imperfectness of the spectral model and/or difference in the details of the
spectral approximation.

Distinction should be made between classes of objects and multiple observations
of individual objects. In the case of luminous BH binaries, spectral variability
of individual sources obeys the general $\Omega$-$\Gamma$ correlation obtained
for BH binaries as class. In the case of Seyfert galaxies, the correlations
holds for a class of objects but can be violated for repeated observations of
individual objects, remaining, however, within the spread of the global
correlation. In our opinion, these results are still inconclusive and require
further investigation as a number of complications are involved in the case of
Seyfert galaxies in comparison with X-ray binaries. The most obvious among those
are lower statistics due to significantly lower brightness of Seyferts and
existence of molecular tori and  broad line regions, which can give additional
contribution to the reflected component, and uncorrelated with the Comptonized
emission on short time scales.

The physical interpretion of the $\Omega$-$\Gamma$ and $\Gamma$-$F$ correlations
will advance our understanding of the geometry of the accretion flow in X-ray
binaries and Seyfert galaxies and impose valuable constrains on the theoretical
models. At present, the correlations appear to be a natural consequence of
co-existence of cold media (e.g., an accretion disc) and a hot Comptonizing
cloud in the vicinity of the BH. Their geometrical closeness results in double
feedback between the two components of the accretion flow. The cold medium
provides seed soft photons for the Comptonization as well as it reprocesses and
reflects the hard radiation from the hot plasma. If significant fraction of soft
seed photons is due to reprocessing of the hard radiation of the hot cloud, the
more reprocessing results in more cooling of the hot plasma and,
correspondingly, in the softer X-ray spectra. Among the several specific
geometries proposed, the most promising appears to be the disc-spheroid  model
with variable overlap between the hot and cold components of the accretion flow.

Since the characteristic temperature of the accretion disk in AGNs is much lower
than in BH binaries, a generic prediction of this type of models is that for a
given value of reflection the value of the spectral index will be higher in AGNs
than in BH binaries, This prediction is in good agreement with the observed
behavior.

We presented a diagnostic to test an interpretation of the $\Omega$-$\Gamma$
correlation as due to strong ionization of the disc surface layer. It utilizes
the independence of the Klein-Nishina cutoff in the reflected spectrum of the
ionization state. Applied to existing data, this diagnostic does not support
that interpretation.

We found that the pattern of broad-band spectral variability of Seyfert galaxies
on day-to-month time scale includes a pivoting of a power law spectrum with the
pivot at a high energy, usuallly above a few hundred keV. The pivoting well
explains the linear correlations between the logarithm of the X-ray flux and
$\Gamma$ seen in both BH binaries and Seyferts. Then, the observed high pivot
energies rule out the interpretation of the $\Omega$-$\Gamma$ correlation is
Seyferts as due to a time-lag effect.

We discuss also the correlation between X-ray and radio fluxes in BH binaries
and its physical implications. We conclude that this correlation is most likely
due to relation between the level of X-ray emission and the rate of ejection of
radio-emitting clouds forming a compact jet. Although peculiar sources might
exist, it seems highly unlikely that the correlation is due to synchrotron
origin of the X-ray emission from BH binaries in general.

\section*{ACKNOWLEDGMENTS}

This research has been supported by grants from KBN (5P03D00821, 2P03C00619p1,2)
and the Foundation for Polish Science. We thank J. Poutanen, P.-O. Petrucci, J.
Malzac, H. Falcke, M. Sikora, R. Moderski, S. Markoff and K. Leighly for
valuable discussions and/or comments. We also acknowledge the referee for
important comments on the original version of this work.

\appendix
\section{Pivoting}

Let us consider spectral variability consisting of pivoting. Namely, an initial
spectrum (a power law or not) is multiplied by $(E/\ep)^{\delta}\equiv f$, where
$\ep$ is the pivot energy and $\delta$ is a perturbation spectral index. In
particular, when the initial spectrum is a power law, we have a variable
power-law photon spectrum, ${\rm d}{\dot N}/ {\rm d}E = C (E/\ep
)^{-\Gamma}$, where $C$ is a constant. Then there is linear relation between the
logarithm of the monochromatic flux at a given energy, $E$, and the variable
slope,
\begin{equation}
\ln({\rm d}{\dot N}/ {\rm d}E) = \ln(\ep/E) \Gamma  + \ln C.
\end{equation}
A similar linear relation holds approximately for the energy flux in an interval
from $E_1$ to $E_2$. Thus, pivoting results in an (approximate) linear
dependence between the logarithm of the flux (not flux itself) and the spectral
index, as illustrated by the dashed lines in Fig.\ \ref{fg}.

\subsection{Moments of the flux}

If pivoting variability occurs on time scales shorther than a given
observation, the measured average spectrum will consist of the initial spectrum
(in particular, a power law itself) times the average of $f$, $\bar f$. If
$\ep=$ constant and the distribution in time of $\delta$ is uniform from
$-\Delta_\Gamma$ to $+ \Delta_\Gamma$, the flux average is given by,
\begin{equation}
{\bar f}={\sinh x\over x}=1+{x^2 \over 6} + O(x^4),
\end{equation}
where
\begin{equation}
x\equiv \Delta_\Gamma\ln(E/\ep).
\end{equation}
If the distribution of $\delta$ is Gaussian, i.e., $\propto
\exp[-(\delta/\Delta_\Gamma)^2/2]$ (where now $\Delta_\Gamma$ is the standard
deviation of
the distribution of $\delta$), we have,
\begin{equation}
{\bar f}={\rm e}^{x^2/2}=1+{x^2 \over 2} + O(x^4).
\end{equation}
Note that the concave form of these spectra yield strong departures from the
initial power law at $E\ll \ep$. Pivoting on short time scales can account (at
least in part) for the soft X-ray excesses commonly seen in Seyferts and BH
binaries. Pivoting may also explain an apparently concave part of the
extragalactic \g-ray spectrum (Stecker \& Salamon 1996), although it requires
the pivot energy to be within that part (which issue was not considered by those
authors).

This variability will also contribute to the flux variance, $\sigma^2= \langle
(f-{\bar f})^2\rangle$. The variance normalized to the average flux is given by
\begin{equation}
{\sigma^2\over {\bar f}^2}=
x \coth x-1 ={x^2 \over 3} + O(x^4),
\label{sigma_u}
\end{equation}
for the uniform distribution of $\delta$, and
\begin{equation}
\sigma^2={\rm e}^{2x^2}-{\rm e}^{x^2},\quad {\sigma^2\over {\bar f}^2}={\rm
e}^{x^2}-1=x^2 +O(x^4),
\label{sigma_g}
\end{equation}
for the Gaussian distribution of $\delta$. With addition of some
energy-independent variability, $\sigma^2_0$, the ratio $(\sigma^2+ \sigma^2_0)/
{\bar f}^2$ reproduces well the fractional variability of Cyg X-1  in the hard
state on long time scales, see Fig.\ \ref{cyg_rms}.

Then, the skewness, $s=\langle (f-{\bar f})^3\rangle /\sigma^3$, is
given by,
 \begin{eqnarray}
\lefteqn{ s =
\frac{6 + x\left[ x + 3
        \left(x\coth x-3 \right)\coth x  \right] }
    {3\left(  x\coth x-1 \right)^{3/2}} } \nonumber \\
\lefteqn{
\quad\,\,\, ={2\sqrt{3} \vert x\vert \over 5} + O(x^3),}
 \end{eqnarray}
for the uniform distribution of $\delta$, and
\begin{equation}
s=\left({\rm e}^{x^2} -1\right)^{1/2}\left({\rm e}^{ x^2}+2\right)
=3\vert x\vert  + O(x^3),
\label{skew_g}
\end{equation}
for the Gaussian distribution of $\delta$. This quantity may be useful for
testing whether a given variability pattern is related to pivoting. Also, the
kurtosis, $S=\langle (f-{\bar f})^4\rangle /\sigma^4-3$, is given by,
\begin{equation}
S={\rm e}^{2x^2}
\left[{\rm e}^{x^2}\left({\rm e}^{x^2}+2\right)+3\right]-6
=16x^2 + O(x^4),
\label{K_g}
\end{equation}
for the Gaussian distribution of $\delta$.

On the other hand, it is rather unlikely that the pivot energy is completely
constant in an astrophysical system. Instead, a range of pivoting energy is
expected from some physical constraints on the emitting plasma, see fig.\
3 in ZG01 and fig.\ 14 in Z02. Therefore, we also consider
a case when the pivot energy is distributed log-normally, i.e., given by $\ln\ep
=\ln \ep^0 -\epsilon$, where $\epsilon$ is distributed $\propto
\exp[-(\epsilon/\Delta_{\rm p})^2/2]$ and $\Delta_{\rm p}$ is the standard
deviation of this distribution. When $\delta$ is also distributed normally, the
average departure from the original spectrum is given by,
\begin{eqnarray}
\lefteqn{
{\bar f}={{\rm e}^{x^2/[2(1-\Delta^2)]}\over (1-\Delta^2)^{1/2}} }
\nonumber \\
\lefteqn{
\quad = {1\over (1-\Delta^2)^{1/2}}+{x^2 \over 2(1-\Delta^2)^{3/2}} + O(x^4), }
\end{eqnarray}
where
\begin{equation}
\Delta \equiv \Delta_\Gamma \Delta_{\rm p}<1,\quad x\equiv
\Delta_\Gamma\ln(E/\ep^0).
\end{equation}

The variance is then,
\begin{equation}
\sigma^2={{\rm e}^{2x^2/(1-4\Delta^2)}\over (1-4\Delta^2)^{1/2}}
-{{\rm e}^{x^2/(1-2\Delta^2)}\over (1-2\Delta^2)^{1/2}},
\end{equation}
and the variance normalized to the average flux is,
\begin{eqnarray}
\lefteqn{
{\sigma^2\over {\bar f}^2}= {1-\Delta^2\over
\left(1-4\Delta^2\right)^{1/2} }\exp\left[ (1+2\Delta^2)x^2\over
(1-4\Delta^2)(1-\Delta^2) \right]+} \nonumber \\
\lefteqn{\qquad\quad
- {1-\Delta^2\over \left(1-2\Delta^2\right)^{1/2} }\exp\left[ \Delta^2 x^2\over
(1-2\Delta^2)(1-\Delta^2) \right] } \nonumber \\
\lefteqn{\quad\,\,
=(1-\Delta^2)\left[ \left(1-4\Delta^2\right)^{-1/2} -
\left(1-2\Delta^2\right)^{-1/2} \right] + } \nonumber \\
\lefteqn{\qquad\quad
\left[ {1+2\Delta^2 \over \left(1-4\Delta^2\right)^{3/2}} - {\Delta^2\over
\left(1-2\Delta^2\right)^{3/2} }\right] x^2 + O(x^4), }
\label{sigma_f_2}
\end{eqnarray}
for $\Delta<1/2$. Note that a finite range of the pivot energy results in a
finite variance at $E=\ep^0$, as given by the third line of equation
(\ref{sigma_f_2}), in contrast to the case with the fixed pivot, equation
(\ref{sigma_g}). Thus, the above dependence fits well the long-term
behaviour of Cyg X-1 in the hard state without an additional constant, see Fig.\
\ref{cyg_rms}.

The skewness  is given by,
\begin{eqnarray}
\lefteqn{
\langle (f-{\bar f})^3\rangle = s\sigma^3=}
\nonumber \\
\lefteqn{\quad\,
{{\rm e}^{9x^2/(2-18\Delta^2)}\over (1-9\Delta^2)^{1/2}}
+{2{\rm e}^{3x^2/(2-6\Delta^2)}\over (1-3\Delta^2)^{1/2}}-{3{\rm
e}^{5x^2/(2-10\Delta^2)}\over (1-5\Delta^2)^{1/2}},
}
\label{skew2}
\end{eqnarray}
for $\Delta<1/3$.

Above, we assumed that variations of the pivot energy are uncorrelated with
those of the spectral index. Note that a correlation between these quantities
would result in time lags between light curves measured at different energies
(Kotov, Churazov \& Gilfanov 2001).

\subsection{Moments of the logarithm of flux}

The moments of the distributions above assume especially simple forms for the
logarithm of flux. We define,
\begin{equation}
g\equiv \ln f= \delta \ln (E/\ep),
\end{equation}
where $f$ is a departure from the initial spectrum, as above.
Then, for a Gaussian distribution of $\delta$ and a
constant $\ep$, the distribution of $g$ is Gaussian itself, with
\begin{equation}
{\bar g}=0,\quad \sigma^2=x^2,\quad s=0,\quad S=0.
\end{equation}
Thus, the apparently non-Gaussian distribution of the linear flux (equations
[\ref{skew_g}]--[\ref{K_g}]) becomes completely Gaussian when its logarithm is
used. Such log-normal distributions are, in fact, very common in natural
phenomena ranging from terrestrial lightning to \g-ray bursts (McBreen et al.\
1994). Note then that a positive skewness of the distribution of a linear
variable (e.g., Leighly 1999) is not necessary an indication of non-Gaussianity
of the underlying processes.

In the case of Gaussian distributions of both $\delta$ and $\epsilon$,
$g$ is a product of two variables ($\delta$ and $\epsilon+{\rm constant}$),
each having  a Gaussian distribution, and
\begin{equation}
{\bar g}=0,\quad\! \sigma^2=\Delta^2+x^2,\quad\! s=0,\quad\!
S={6\Delta^2(\Delta^2+2x^2)\over (\Delta^2+x^2)^2}.
\end{equation}
Note that $\sigma^2$ and $S$ are not simply products of the corresponding
quantities for each of the two variables due to a non-zero center value of the
latter. Then, the resulting distribution of $g$ is not completely Gaussian, and,
in particular, $S>0$.

\bsp
\label{lastpage}
\end{document}